**Deep phenotyping of cardiac function in heart transplant patients using cardiovascular systems models**

Amanda L. Colunga*, Karam G. Kim*, N. Payton Woodall*, Todd F. Dardas, John H. Gennari, Mette S. Olufsen, Brian E. Carlson

* These authors contributed equally to this manuscript

## Abstract

Heart transplant patients are followed with periodic right heart catheterizations (RHCs) to identify post-transplant complications and guide treatment. Post-transplant positive outcomes are associated with a steady reduction of right ventricular and pulmonary arterial pressures, toward normal levels of right-side pressure (about 20mmHg) measured by RHC. This study shows more information about patient progression is obtained by combining standard RHC measures with mechanistic computational cardiovascular systems models. The purpose of this study is two-fold: to understand how cardiovascular system models can be used to represent a patient's cardiovascular state, and to use these models to track post-transplant recovery and outcome. To obtain reliable parameter estimates comparable within and across datasets, we use sensitivity analysis, parameter subset selection, and optimization to determine patient specific mechanistic parameter that can be reliably extracted from the RHC data. Patient-specific models are identified for ten patients from their first post-transplant RHC and longitudinal analysis is done for five patients. Results of sensitivity analysis and subset selection show we can reliably estimate seven non-measurable quantities including ventricular diastolic relaxation, systemic resistance, pulmonary venous elastance, pulmonary resistance, pulmonary arterial elastance, pulmonary valve resistance and systemic arterial elastance. Changes in parameters and predicted cardiovascular function post-transplant are used to evaluate cardiovascular state during recovery in five patients. Of these five patients, only one patient showed inconsistent trends during recovery in ventricular pressure-volume relationships and power output. At the four-year post-transplant time point this patient exhibited biventricular failure along with graft dysfunction while the remaining four exhibited no cardiovascular complications.

## Introduction

Diagnosis and treatment plans for patients with cardiovascular pathophysiologies are currently being guided with an increasing number of noninvasive and minimally invasive clinical measures by statistically inferring correlations between measurements and diagnosis/treatments. However, this approach is limited in scope since the underlying physiological mechanisms associated with the success or failure of a given treatment for a given patient cannot be discriminated. One example is the assessment of cardiovascular function using right heart catheterization measures after heart transplants. For this patient group, repeated right heart catheterization (RHC) measurements of ventricular and pulmonary arterial pressure are used to monitor post-transplant pulmonary hypertension, which if not resolved, can lead to complications in post-transplant recovery (Greenberg *et al.*, 1985; Bhatia *et al.*, 1987; Young *et al.*, 1987; Goland *et al.*, 2007). Traditionally, RHC measurements are used to inform post-transplant treatment and intervention and a close monitoring of these measures has been associated with better outcome. However, these RHC measures only describe the upper-level phenotype of the cardiovascular system and do not explicitly take advantage of the relationships between pressures, volumes, and flow governed by the known physiology of the cardiovascular system. It has been suggested (Armitage *et al.*, 1987; Stobierska-Dzierzek *et al.*, 2001), that a complete evaluation of the cardiovascular system could improve detection and treatment of dysfunction in the transplanted heart. The study presented here builds a patient specific computational methodology (Figure 1) integrating clinical measures and computing time-varying patient specific pressures, flows, and volumes, while estimating mechanistic parameters, which can be incorporated in clinical analysis to guide treatment and assess recovery of heart transplant patients. Many of the estimated mechanistic parameters and predicted cardiovascular variables obtained with the computational approach presented here cannot be easily measured in the clinic.

In addition to the RHC measurements, echocardiography, magnetic resonance, and Doppler imaging have been used to track metrics of cardiac function post-transplantation (Sundereswaran *et al.*, 1998; Dandel *et al.*, 2001; Marie *et al.*, 2001; Sun *et al.*, 2005). These non-invasive modalities are always coupled with RHC measurements in standard clinical protocols and may not on their own provide sufficient information needed to improve diagnosis and treatment protocols. The study by Dandel et al. (2001) utilized tissue Doppler, to determine the optimal times that RHC measurements should be made during recovery, while the other studies focused on an identifying a single biomarker or sets of biomarkers from echocardiography which were used to identify dysfunction and guide therapy. This latter approach is problematic in two ways. First, if a prospective biomarker or set of biomarkers does not discriminate outcomes another biomarker or set of biomarkers must be selected, and the process repeated. Second, this approach ignores the fact that post-transplant recovery and outcome are multifactorial, involving the function of the entire cardiovascular system working in conjunction with the transplanted heart and thus biomarkers focusing only on the transplanted heart have a reduced chance of discriminating patient outcome.

To gain more insight into post-transplant recovery of cardiovascular function, our approach uses mathematical models to analyze RHC deidentified data from patient electronic health records (EHRs) that also contain post-transplant clinical outcome. More specifically, we employ a mechanistic representation of each patient at a specific recovery timepoint using a mathematical model of the cardiovascular system similar to several previously developed models (Smith *et al.*, 2004; Lumens *et al.*, 2009; Beard *et al.*, 2013; Williams *et al.*, 2014). We personalize the model by calculating initial estimates of model parameters using information extracted from the EHR. Next, we use rigorous model analysis techniques to identify and estimate parameters minimizing the least squares error between the model predictions and data.

The final step involves running forward simulations with the personalized instance of the model predicting patient-specific dynamics that can be used to inform the clinical diagnosis and treatment procedure. A vital part of this analysis is to select the right granularity of the model informed by the clinical data, and then quantify which parameters can be estimated given the model and available RHC data. At this point the RHC data used here does not contain any pressure waveforms and is not combined with any direct measures of left ventricular function.

The patient-specific instantiations of the mathematical model are used to infer underlying differences between 10 patients, and then associate how changes in cardiovascular function are related to clinical outcome. By integrating our physiological knowledge of the cardiovascular system with patient-specific data, we are constraining the system to represent the cardiovascular state of an individual patient. In addition, since this is a retrospective analysis, our approach opens the potential to search for early indicators of positive and negative heart transplant outcomes.

Of the 10 patients' datasets obtained, five contain longitudinal RHC measurements at an additional 3-7 post-transplant time points over the span of up to 13 months. We analyze these longitudinal RHC measures from these patients to quantify how underlying cardiovascular function for each of these five patients is changing during post-transplant recovery. Finally, the trends in the predictions of pressure-volume loops and ventricular power output for the right and left ventricle are associated with clinical outcome in each patient.

Table 1. Right heart catheter measurements extracted from the clinical data repository at the University of Washington Medicine Regional Heart Center.

| Patient | $P_{rv}$ systole | $P_{rv}$ diastole | $P_{pa}$ systole | $P_{pa}$ diastole | $P_{pcw}$ average | $P_{sa}$ systole | $P_{sa}$ diastole | CO** [L/min] | HR** [bpm] | Weight [kg] | Height [cm] | Sex |
|---|---|---|---|---|---|---|---|---|---|---|---|---|
| 060 | 36 | 2 | 36* | 14 | 16 | 132 | 79 | 6.3 | 69 | 62 | 180 | M |
| 066 | 28 | 1 | 26 | 13 | 8 | 134 | 79 | 7.8 | 81 | 120 | 194 | M |
| 233 | 40 | 4 | 35 | 19 | 19 | 149 | 83 | 10.3 | 83 | 96 | 172 | F |
| $266^{(1)}$ | 37 | 10 | 34 | 14 | 11 | 154 | 81 | 6.0 | 90 | 75 | 169 | M |
| $266^{(2)}$ | 30 | 6 | 27 | 10 | 9 | 121 | 75 | 6.5 | 94 | 71 | 169 | M |
| $266^{(3)}$ | 29 | 9 | 30* | 13 | 12 | 115 | 82 | 6.1 | 96 | 73 | 169 | M |
| $266^{(4)}$ | 27 | 8 | 25 | 10 | 9 | 112 | 69 | 5.2 | 92 | 73 | 169 | M |
| $266^{(5)}$ | 23 | 5 | 26* | 10 | 8 | 106 | 77 | 5.4 | 94 | 75 | 169 | M |
| $266^{(6)}$ | 27 | 9 | 28* | 12 | 11 | 119 | 81 | 5.9 | 101 | 79 | 169 | M |
| $266^{(7)}$ | 22 | 5 | 22* | 8 | 7 | 107 | 76 | 6.0 | 102 | 75 | 169 | M |
| $363^{(1)}$ | 30 | 7 | 29* | 20 | 12 | 108 | 78 | 4.8 | 109 | 80 | 162 | M |
| $363^{(2)}$ | 22 | 1 | 21 | 13 | 6 | 119 | 87 | 4.1 | 92 | 75 | 162 | M |
| $363^{(3)}$ | 25 | 2 | 25* | 17 | 10 | 113 | 73 | 3.4 | 108 | 79 | 162 | M |
| $363^{(4)}$ | 26 | 1 | 24 | 12 | 7 | 107 | 85 | 4.8 | 119 | 79 | 163 | M |
| $456^{(1)}$ | 44 | 4 | 35 | 13 | 16 | 104 | 67 | 5.5 | 90 | 78 | 170 | M |
| $456^{(2)}$ | 42 | 4 | 36 | 14 | 15 | 122 | 75 | 5.0 | 86 | 78 | 170 | M |
| $456^{(3)}$ | 39 | 2 | 33 | 14 | 13 | 128 | 88 | 4.6 | 93 | 75 | 170 | M |
| $456^{(4)}$ | 23 | 5 | 24* | 13 | 6 | 130 | 79 | 5.0 | 108 | 78 | 170 | M |
| $456^{(5)}$ | 37 | 11 | 33 | 18 | 15 | 118 | 78 | 5.9 | 93 | 78 | 170 | M |
| $456^{(6)}$ | 34 | 12 | 31 | 18 | 14 | 133 | 75 | 5.5 | 103 | 73 | 170 | M |
| $456^{(7)}$ | 30 | 4 | 24 | 13 | 9 | 110 | 74 | 4.7 | 99 | 78 | 170 | M |
| $456^{(8)}$ | 28 | 1 | 24 | 9 | 6 | 99 | 63 | 6.9 | 100 | 74 | 170 | M |
| $558^{(1)}$ | 31 | 3 | 32* | 15 | 14 | 114 | 49 | 7.7 | 90 | 101 | 170 | F |
| $558^{(2)}$ | 38 | 7 | 33 | 12 | 15 | 124 | 77 | 6.1 | 63 | 101 | 170 | F |
| $558^{(3)}$ | 34 | 5 | 33 | 14 | 18 | 144 | 85 | 7.9 | 71 | 101 | 170 | F |
| $558^{(4)}$ | 36 | 1 | 35 | 12 | 16 | 152 | 88 | 7.8 | 73 | 103 | 170 | F |
| $558^{(5)}$ | 45 | 5 | 44 | 21 | 25 | 125 | 71 | 7.8 | 70 | 110 | 170 | F |
| $558^{(6)}$ | 28 | 1 | 27 | 10 | 12 | 128 | 79 | 7.4 | 69 | 103 | 170 | F |
| $558^{(7)}$ | 33 | 2 | 34* | 17 | 12 | 166 | 102 | 7.4 | 69 | 103 | 170 | F |
| $572^{(1)}$ | 31 | 5 | 30 | 12 | 12 | 116 | 66 | 6.4 | 100 | 82 | 180 | M |
| $572^{(2)}$ | 29 | 5 | 24 | 7 | 9 | 119 | 72 | 7.3 | 97 | 82 | 180 | M |
| $572^{(3)}$ | 26 | 2 | 24 | 8 | 7 | 112 | 72 | 6.7 | 100 | 82 | 180 | M |
| $572^{(4)}$ | 26 | 2 | 23 | 10 | 9 | 109 | 71 | 6.5 | 95 | 82 | 180 | M |
| $572^{(5)}$ | 28 | 1 | 24 | 10 | 9 | 112 | 73 | 6.9 | 90 | 82 | 180 | M |
| $572^{(6)}$ | 17 | 4 | 20* | 10 | 10 | 108 | 67 | 6.2 | 88 | 83 | 180 | M |
| $572^{(7)}$ | 22 | 1 | 16 | 9 | 5 | 122 | 71 | 7.1 | 98 | 82 | 180 | M |
| $572^{(8)}$ | 19 | 4 | 17 | 9 | 7 | 115 | 73 | 7.2 | 97 | 82 | 180 | M |
| 794 | 38 | 6 | 39* | 17 | 21 | 98 | 57 | 6.8 | 84 | 44 | 174 | F |
| 839 | 25 | 1 | 18 | 9 | 7 | 135 | 84 | 6.1 | 91 | 83 | 170 | M |

All pressures are given in mmHg. * indicates that when nominal values were calculated pulmonary artery pressure was adjusted to be 95% of the pressure in the right ventricle to enforce a pressure drop in the direction of flow. **Measurements of cardiac output (CO) and heart rate (HR) are given in conventional units [L/min] and [bpm], but to keep consistency of units, in the model these are converted to [mL/s] and [bps]. Subscripts: *rv* – right ventricle, *pa* – pulmonary arteries, *pcw* – pulmonary capillary wedge, *sa* – systemic arteries

# Methods

***RHC measures:*** This study analyzes 10 RHC records from heart transplant patients extracted from the clinical data repository at the University of Washington Medicine Regional Heart Center. This retrospective data capture was approved by the Institutional Review Board (IRB) at the University of Washington. The data in the repository was exported from the Mac-Lab Hemodynamic Recording System (GE Healthcare, Chicago, IL, USA) used in the UW cardiac catheterization lab. The repository was queried for RHC datasets from heart transplant patients with catheterization procedures performed between March 6, 2014 and March 21, 2016. Ten patient records were retrieved, of these, five records contained multiple RHC measures from a four- to twelve months period immediately following the transplant. The datasets contained twelve clinically measured values including: systolic and diastolic pressure measured in the right ventricle, pulmonary artery, and aorta; an average pulmonary capillary wedge pressure, heart rate, cardiac output, body weight, height, and sex, as shown in Table 1.

***Mathematical model:*** Inspired by a previously published cardiovascular systems-level model (Smith *et al.*, 2004), we developed the mechanistic model used here (schematic shown in Figure 2) to study patient-specific cardiovascular function in heart transplant patients. The objective of this study is to create a simple model that can simulate the RHC and systemic arterial blood pressure data. The main differences between our model and the model by Smith et al. are that (1) we ignored the right ventricular and left ventricular pressures and volumes influence on each other via the septal wall otherwise known as ventricular-ventricular interaction (VVI) and (2) we omitted the influence of the inertance of blood flowing through the four heart valves. We justified the former using simple sensitivity analysis by doubling and halving parameters involved with VVI around nominal values for several of our patients. Low sensitivity of the parameters involved with ventricular-ventricular interaction did not justify the significant increase in model complexity. This observation agrees with findings in the literature, which note that VVI mainly impact left ventricular dynamics for patients with very high right ventricular pressures and volumes associated severe pulmonary hypertension (Maughan *et al.*, 1981; Gan *et al.*, 2006). Inertance of blood flowing through the four valves in the heart could be of importance for prediction of waveforms (especially the aortic pressure waveform). However, the RHC measures used here are not given as waveforms in the EHR but simply are provided as maximal and minimal pressures in systole and diastole respectively. Therefore, this RHC data from the EHRs do not contain the information needed to identify inertance parameters.

The cardiovascular systems model depicted in Figure 2 is used to predict blood pressure, flow, and volume in the heart's left and right ventricles, and the pulmonary and systemic arteries and veins. The model is analogous to several RC electrical components in series where current is analogous to blood flow $Q$ [mL/s], voltage to pressure $P$ [mmHg], and charge to volume $V$[mL]. Note, the blood flow $Q$ is directly related to cardiac output which is the average flow over the period, $T$, of the cardiac cycle, i.e. $\mathrm{CO} = \frac{1}{T}\int_0^T Q\, dt$ [mL/s]. In addition, elastance $E$ [mmHg/mL] is reciprocal of compliance, which is analogous to capacitance, while $R$ [mmHg·s/mL] represents the resistance to flow within a compartment. Diodes are used to simulate one-way valves preventing blood from exiting a ventricle through a closed aortic, mitral, pulmonary or tricuspid heart valve. The muscle contractions within the heart are modeled using a Gaussian activation function defined over one cardiac cycle as

$$A(\tilde{t}) = e^{-a(\tilde{t}-T/2)^2}, \tag{1}$$

where $a$ is a scaling factor with the value of heart rate from Table 1 and units of $1/\mathrm{s}^2$ to make the exponent unitless, $T = 1/H$ [s], where $H$ is heart rate in beats per second, is the period of the cardiac cycle and $\tilde{t} = mod(t, T)$ [s] is the time from the start of the current cardiac cycle. This relation creates a symmetric curve about $T/2$ bounded by $0 \leq A(\tilde{t}) \leq 1$ and denotes the relative contribution of the systolic and diastolic pressure development during the cardiac cycle. The end-systolic pressure $P_{es}$ and volume $V_{es}$ in the left and right ventricles are assumed to be linearly related to the end-systolic ventricular elastance $E_{es}$ via

$$P_{es} = E_{es}(V_{es} - V_d), \tag{2}$$

where $V_d$ is the end-systolic volume at zero pressure. In addition, the end-diastolic pressure $P_{ed}$ is related nonlinearly to the end-diastolic volume $V_{ed}$ by

$$P_{ed} = P_0\left[e^{\lambda(V_{ed}-V_0)} - 1\right], \tag{3}$$

where $P_0$ is the pressure at the unstressed volume $V_0$, and $\lambda$ [mL$^{-1}$] specifies the steepness of the exponential pressure-volume relationship in diastole. The pressure in the left ventricle at any time in the cardiac cycle is calculated by combining Eq. (2) and (3) giving

$$P(t) = A(\tilde{t})P_{es} + [1 - A(\tilde{t})]P_{ed} + P_{th}, \tag{4}$$

where $P_{th} = 0$ is the assumed tissue pressure in the thoracic cavity. As $A(\tilde{t})$ changes throughout the cardiac cycle, Eq. (4) shifts the contributions of the systolic and diastolic pressure terms to give the total pressure in the ventricle. When $A(\tilde{t})$ is equal to 1 the pressure is described solely by eq. (2) and when equal to 0 by the diastolic pressure-volume curve of eq. (3). Since the pulmonary arteries and veins are located in the thoracic cavity, the pressure and volume are related as

$$P = EV + P_{th}, \tag{5}$$

while the majority of the systemic arteries and veins are outside the thoracic cavity where

$$P = EV. \tag{6}$$

Blood flow in and out of each compartment is proportional to the difference between the compartments input and output pressures and represented by the fluid equivalent to Ohm's Law

$$Q = \frac{P_{in} - P_{out}}{R} \tag{7}$$

and conservation of volume, where the change in the compartmental volume must equal the difference in flow in and flow out, implying that

$$\frac{dV}{dt} = Q_{in} - Q_{out} \tag{8}$$

for each compartment. To model the one-way heart valves, we set flow to zero when pressure across the valve indicates the valve is closed, giving

$$Q_{valve} = \begin{cases} \dfrac{P_{in} - P_{out}}{R} & \text{if } P_{in} > P_{out} \\ 0 & \text{otherwise} \end{cases} \tag{9}$$

for the aortic ($av$), tricuspid ($tc$), pulmonary ($pv$), and mitral ($mt$) valves. A list of equations making up the model is given in Appendix A, and a version of the model implemented in MATLAB (The MathWorks, Inc., Natick, MA) is available at github.com/alcolunga/Heart_Tx_CVS_Model and at https://wp.math.ncsu.edu/cdg/publications/. In summary, the system of equations can be written in the form

$$\frac{dx}{dt} = f(x, t\colon \theta)$$

$$y = g(x, t\colon \theta)$$

$$x = \{V_{lv}, V_{sa}, V_{sv}, V_{rv}, V_{pa}, V_{pu}\}$$

$$\theta = \{E_{lv}, V_{d,lv}, P_{0,lv}, \lambda_{lv}, V_{0,lv}, E_{rv}, V_{d,rv}, P_{0,rv}, \lambda_{rv}, V_{0,rv},$$

$$E_{pa}, E_{pu}, R_{pul}, P_{th}, E_{sa}, E_{sv}, R_{sys}, R_{mt}, R_{av}, R_{tc}, R_{pv}\}$$

$$y = \{P_{rv}, P_{pa}, P_{pu}, P_{sa}, Q\}.$$

Where $x$ are the state variables of the model, $t$ is time, $\theta$ are the model parameters and $y$ are the model outputs, which we take the max of, min of or average in order to compare with our clinical data. The data is expected to match model output as

$$y_i^d = y(t_i) + \epsilon_i, \qquad i = 1, \ldots, K,$$

where $K$ denotes the number of time-points, and $\varepsilon_i$ is the error assumed to be independent identically distributed random variables with mean $E[\varepsilon_i] = 0,$ covariance $\text{cov}(\varepsilon_i, \varepsilon_j) = 1$, and constant variance $var(\varepsilon_i) = \mu^2$. The equations are solved under the assumptions that the cardiac cycle is initialized to be at end-diastole, i.e. the volumes in the left and right heart are at their maximum, while the venous and arterial volumes are initialized at their average values.

***Nominal parameter values***: Nominal parameter values for each patient RHC are determined from the clinical data and literature values using an approach similar to a previous method for the rat cardiovascular system (Marquis *et al.*, 2018) but translated to the human. The nominal parameters and how they are obtained are shown in Table 2.

Blood volume: In order to compute nominal values of all elastances, the blood volume distribution in the cardiovascular system must be estimated. Total blood volume (TBV in mL) (Nadler *et al.*, 1962) and body surface area (BSA in m$^2$) (Du Bois & Du Bois, 1916) are calculated as functions of height (Hgt in cm), weight (BW in kg), and sex as

$$\text{TBV} = \begin{cases} (0.3669\text{Hgt}^3 + 0.03219\ \text{BW} + 0.641)\ 1000 & \text{for males} \\ (0.3561\ \text{Hgt}^3 + 0.03308\ \text{BW} + 0.1833)\ 1000 & \text{for females} \end{cases} \tag{10}$$

$$\text{BSA} = \begin{cases} 0.0005795\ \text{BW}^{0.38}\ \text{Hgt}^{1.24} & \text{for males} \\ 0.0009755\ \text{BW}^{0.46}\ \text{Hgt}^{1.08} & \text{for females.} \end{cases} \tag{11}$$

TBV is used to estimate the volumes in the systemic vasculature while BSA is used to estimate mean left and right ventricular volumes as shown later. Following approximations by Beneken et. al. (1967) the total blood volume is distributed as 3.5% in the left ventricle, 3.5% in the right ventricle, 3% in the pulmonary arteries, 11% in the pulmonary veins, 13% in the systemic arteries, and 64% in the systemic veins. The compartment models only track stressed volume which we assume are approximately the following for the distributed volumes in each individual

Table 2. Equations for calculating nominal parameter values.

| | Parameter | Units | Equation | Mean ± std | Reference |
|---|---|---|---|---|---|
| Heart Parameters | $H$ | $\mathrm{s}^{-1}$ | heart rate | 90.66 ± 12.91 | |
| | $T$ | s | $H^{-1}$ | 0.33 ± 0.05 | |
| Left Ventricle | $E_{lv}$ | $\frac{\text{mmHg}}{\text{mL}}$ | $\frac{P_{sa,syst} - P_{th}}{V_{lv,m} - V_{d,lvf}}$ | 1.7 ± 0.6 | * |
| | $V_{d,lv}$ | mL | | 10 | (Williams, 2014) |
| | $P_{0,lv}$ | mmHg | | 0.125 | ** |
| | $\lambda_{lv}$ | $\mathrm{mL}^{-1}$ | | 0.029 ± 0.005 | ** |
| | $V_{0,lv}$ | mL | | 12 | ** |
| Right Ventricle | $E_{rv}$ | $\frac{\text{mmHg}}{\text{mL}}$ | $\frac{P_{pa,syst} - P_{th}}{V_{rv,m} - V_{d,rvf}}$ | 0.44 ± 0.23 | * |
| | $V_{d,rv}$ | mL | | 9 | ** |
| | $P_{0,rv}$ | mmHg | | 0.25 | ** |
| | $\lambda_{rv}$ | $\mathrm{mL}^{-1}$ | | 0.024 ± 0.004 | ** |
| | $V_{0.rv}$ | mL | | 10.8 | ** |
| Pulmonary Vasculature | $E_{pa}$ | $\frac{\text{mmHg}}{\text{mL}}$ | $\frac{P_{pa,syst} - P_{th}}{\text{CBV}_{pa}}$ | 0.32 ± 0.09 | * |
| | $E_{pu}$ | $\frac{\text{mmHg}}{\text{mL}}$ | $\frac{P_{pcw}}{\text{CBV}_{pu}}$ | 0.019 ± 0.08 | * |
| | $R_{pul}$ | $\frac{\text{mmHg} \cdot \text{s}}{\text{mL}}$ | $\frac{P_{pa,syst} - P_{pcw}}{\text{CO}}$ | 0.16 ± 0.05 | Ohm's Law/Data |
| Systemic Vasculature | $E_{sa}$ | $\frac{\text{mmHg}}{\text{mL}}$ | $\frac{P_{sa,syst}}{\text{CBV}_{sa}}$ | 0.69 ± 0.09 | * |
| | $E_{sv}$ | $\frac{\text{mmHg}}{\text{mL}}$ | $\frac{\frac{1}{3}P_{sv,syst} + \frac{2}{3}P_{sv,diast}}{\text{CBV}_{sv}}$ | 0.02 ± 0.01 | * |
| | $R_{sys}$ | $\frac{\text{mmHg} \cdot \text{s}}{\text{mL}}$ | $\frac{P_{sa,syst} - (\frac{1}{3}P_{sv,syst} + \frac{2}{3}P_{sv,diast})}{\text{CO}}$ | 1.14 ± 0.25 | Ohm's Law/Data |
| Heart Valves | $R_{mt}$ | $\frac{\text{mmHg} \cdot \text{s}}{\text{mL}}$ | $\frac{P_{pu,diast} - P_{lv,diast}}{\text{CO}}$ | 0.0025 ± 0.0009 | Ohm's Law/Data |
| | $R_{av}$ | $\frac{\text{mmHg} \cdot \text{s}}{\text{mL}}$ | $\frac{P_{lv,syst} - P_{sa,syst}}{\text{CO}}$ | 0.029 ± 0.006 | Ohm's Law/Data |
| | $R_{tc}$ | $\frac{\text{mmHg} \cdot \text{s}}{\text{mL}}$ | $\frac{P_{sv,diast} - P_{rv,diast}}{\text{CO}}$ | 0.0011 ± 0.0008 | Ohm's Law/Data |
| | $R_{pv}$ | $\frac{\text{mmHg} \cdot \text{s}}{\text{mL}}$ | $\frac{P_{rv,syst} - P_{pa.syst}}{\text{CO}}$ | 0.028 ± 0.025 | Ohm's Law/Data |

The total flow through the system equals CO (converted to mL/s), CBV denotes the circulating blood volume, *P* the blood pressure, and *H* heart rate (converted to 1/s). Subscripts: *lv* denotes the left ventricle, *rv* the right ventricle, *lvf* the left ventricular free wall, *rvf* the right ventricular free wall, *syst* systole, *diast* diastole, *m* mean, *d* the dead space, *pa* pulmonary arteries, *pu* pulmonary veins, s*a* systemic arteries, and *sv* systemic veins. Entries noted with * are computed by rearranging the model equations and using available data as described in the methods section. Entries noted with ** are set by hand fitting normal CV function model output to the Smith model output using values in the CellML version of the Smith model (Nielsens, 2010) as a guide.

compartment: 27% in the systemic arteries, 58% in the pulmonary arteries, 18% in the systemic veins, and 11% in the pulmonary veins.

Pressure: In order to estimate nominal parameter values for elastances and resistances, pressures must be estimated or extracted from the clinical data. EHRs with RHC procedures include measurements of systolic and diastolic pressures in the right ventricle and the main pulmonary artery along with the capillary wedge pressure which effectively represents the mean pressure in the pulmonary veins. In addition, systemic arterial blood pressure is measured using a pressure cuff. Using these values, we can estimate other pressures in the cardiovascular system. First, we estimate the diastolic pressure in the pulmonary veins. For this, we use the capillary wedge pressure which we assume to represent the mean venous pressure. We then assume that the pulmonary venous pulse pressure is roughly 20% of the measured pulmonary arterial pulse pressure as based on observations in a normotensive human study (Caro *et al.*, 1967) and then calculate the diastolic pressure as the mean pressure minus 33% of the pulse pressure.

$$P_{pu,pp} = 0.2P_{pa,pp}, \tag{12}$$

$$P_{pu,diast} = P_{pcw} - 0.33P_{pu,pp}. \tag{13}$$

In addition, we need to estimate systolic and diastolic pressures for the left ventricle, and the vena cava. The systolic pressure in the left ventricle is of the same order of magnitude as the systolic arterial pressure and the diastolic pressures in the left ventricle and vena cava are similar to the pulmonary venous diastolic pressures and right ventricular diastolic pressure. Assuming a 2.5% pressure drop across the mitral, aortic, and tricuspid valves, we can approximate these three pressures as

$$P_{lv,syst} = 1.025P_{sa,syst}, \tag{14}$$

$$P_{lv,diast} = 0.975 P_{pu,diast}, \quad (15)$$

$$P_{sv,diast} = 1.025 P_{rv,diast}. \quad (16)$$

Furthermore, we estimate the systolic vena cava pressure assuming that the pulse pressure in the vena cava $P_{sv,pp}$ is 5% of systemic arterial pulse pressure, giving

$$P_{sv,syst} = P_{sv,diast} + P_{sv,pp}. \quad (17)$$

<u>*Stroke volume:*</u> Using the cardiac output (CO [mL/s]) and heart rate ($H$ [1/s]) data we can calculate the stroke volume, SV [mL], by

$$\text{SV} = \frac{\text{CO}}{H}. \quad (18)$$

<u>*Volume:*</u> Using a linear regression from Gutgesell and Rembold (1990), we calculate the end-diastolic left ventricular volume as

$$V_{lv} = 93\ \text{BSA} - 16, \quad (19)$$

where BSA denotes the body surface area calculated in Eq. (11). From Eqs. (18) and (19), we calculate the volume of the left ventricle at the end of systole as

$$V_{lv} = V_{lv} - \text{SV}. \quad (20)$$

Accurate measurements of right ventricular volumes are difficult because of the complex geometry of the right ventricular chamber. For the purpose of our nominal parameter calculations we assumed that right ventricular end-diastolic and end-systolic volumes were estimated at roughly 0.9 times the corresponding left ventricular volumes based on observations that the right ventricle chamber volumes are slightly smaller than in the left heart (Hergan *et al.*, 2008; Tamborini *et al.*, 2010).

*Elastance:* Nominal elastance parameter values are needed for all compartments. Nominal elastance parameters are calculated by combining information in Eqs. (2), (5), and (6) with the estimated compartmental blood volume (CBV) in compartment $i$ to give

$$E_i = \frac{P_{i,syst} - P_{th}^*}{CBV_i - V_{d,i}^*}. \tag{21}$$

where $i$ is $sa, sv, rv, pa, pu$ and $lv$. Equation (21) has three forms. For compartments inside the thorax ($pa, pu, rv, lv$), the systolic pressure is offset by the intrathoracic tissue pressure $P_{th}^*$. This term is not included in calculations predicting elastance in the systemic arteries and veins as these mostly are outside the thorax. Finally, the CBV is offset by $V_{d,i}^*$, a dead space volume, in compartments representing the left and right ventricles.

*Resistances:* Nominal values of the resistances in each compartment of the model are calculated from the fluid equivalent of Ohm's Law using measured cardiac output (CO) as our baseline flow and the estimated or measured pressures for $P_{in}$ and $P_{out}$

$$R_i = \frac{P_{in} - P_{out}}{\mathrm{CO}}. \tag{22}$$

In 12 sets of RHC data the systolic pulmonary artery pressure was recorded as greater than or equal to the systolic right ventricular pressure. Since this would yield an unrealistic zero or negative resistance, for these patients we set

$$P_{pa,syst} = 0.95 P_{rv,syst} \tag{23}$$

to estimate the nominal value of the resistance across the pulmonary valve.

***Sensitivity Analysis:*** With unique sets of nominal parameter values determined for each patient RHC we use local sensitivity analysis to determine the relative importance of the parameters in the neighborhood of the nominal values. Given that both the model output and parameters contain quantities of different orders of magnitudes and units, we compute a relative sensitivity matrix $S$ defined by

$$S_{i,j} = \frac{\partial y(t_i, \theta)}{\partial \theta_j} \frac{\theta_j}{y_i^d}, \tag{24}$$

where $y(t_i, \theta)$ represents the model output at time $t_i$, $\theta_j$ denotes the $j$'th parameter, and $y_i^d$ denotes the data measured at time $t_i$. Due to the complexity of the model, $S$ is difficult to calculate analytically, so similar to Pope et al. (2009) we use finite differences to estimate $S$ by

$$S_{i,j} = \frac{y(t_i, \theta + he_j) - y(t_i, \theta)}{h} \frac{\theta_j}{y_i^d}. \tag{25}$$

Here $h = \sqrt{\epsilon}$ where $\epsilon$ is the tolerance (set at $10^{-12}$) of the ODE solver, and $e_j$ is the unit vector in the $j$'th direction. To approximate the influence that each parameter has on the model, we rank the sensitivities, $R_j$, as the two-norm over each column of $S$

$$R_j = \left( \sum_{i=1}^{N} S_{i,j}^2 \right)^{\frac{1}{2}}. \tag{26}$$

***Parameter Subset Selection:*** In addition to being sensitive, it is important that estimated parameters are not correlated (Marquis *et al.*, 2018). To determine a subset of independent parameters we used sensitivity-based covariance analysis (Olufsen & Ottesen, 2013) to get *a priori* insight into potential correlation structure. For all sensitive parameters, we calculate the covariance matrix

$$c_{ij} = \frac{C_{ij}}{\sqrt{C_{ii} C_{jj}}}, \quad C = (S^T S)^{-1}, \tag{27}$$

where $S$ denotes the sensitivity matrix. This formulation is valid under the assumption that the variance is constant. The covariance matric $C$ is defined if $S^T S$ (also known as the Fisher information matrix) can be inverted necessitating the *a priori* removal of parameters that are perfectly correlated. Parameters for which $c_{ij} > \gamma$ (here we let $\gamma = 0.9$) are denoted correlated. Following the structured correlation method by Ottesen and Olufsen (2013), the covariance matrix is analyzed for all sensitive parameters. It is an iterative algorithm that removes the least sensitive parameter from a pair of correlated parameters. Parameters are removed sequentially until an uncorrelated subset is identified. Before the structured analysis discussed above, analytical knowledge is used to identify parameters that appear in structurally correlated combinations. All parameters that are removed from the subset are fixed at their nominal value. This analysis is local in nature as the sensitivity matrix is evaluated at nominal parameter values determined for each patient.

***Model Optimization and Parameter Identifiability***: Next, we use the cardiovascular system model to reproduce the clinical measures from the RHC procedure along with systemic arterial blood pressure. Clinical measures of patient height, weight, and sex are used to estimate the total blood volume (Eq. 10) and heart rate is used to drive the model (Eq. 1). The remaining measures are matched to the output of the model simulation (light gray shaded measures in Table 1). For each patient, we estimate a subset of parameters $\theta^*$, that minimize the least squares error

$$J = r^T r, \tag{28}$$

where $r = \{r_1, r_2, r_3, r_4\}$ is the residual vector containing 8 entries $r_i, 1 \leq i \leq 8$, where

$$r_k^1 = \frac{1}{\sqrt{8}} \frac{\max[P_k(\tilde{t})] - P_{k,syst}^d}{P_{k,syst}^d}, \quad k = \{sa, pa, rv\} \tag{29}$$

$$r_{pcw}^3 = \frac{1}{\sqrt{8}} \frac{\frac{\int_0^T P_{pu}(t_i)\, dt}{T} - P_{pcw}^d}{P_{pcw}^d}, \tag{31}$$

$$r_{sys}^4 = \frac{1}{\sqrt{8}} \frac{\frac{\int_0^T Q(t_i)\, dt}{T} - \mathrm{CO}^d}{\mathrm{CO}^d}, \tag{32}$$

where quantities with superscript $d$ refer to data. For each term values are calculated over one cardiac cycle after the system has reached a steady state of pulsatile pressures and flows over $\tilde{t} \in (0, T)$. The capillary wedge pressure ($P_{pcw}$) and cardiac output ($CO$) represent average values over the cardiac cycle. Therefore, time varying quantities $Q$ and $P_{pu}$ are averaged over the stable cardiac cycle before being compared to data. Given that quantities minimized are of different units, each residual function is divided by a characteristic value for the quantity. Point estimates for the identifiable parameters are obtained using the Levenberg-Marquardt optimization routine by Kelley (1999) and since parameters vary in magnitude we estimate the log-scaled parameters as outlined in Marquis et al. (2018). To ensure convergence we repeated parameter estimation starting with 8 initial parameter sets varying the nominal parameter values by 10%.

To overcome the limitation of the local approach, similar to Marquis et al. (2018) we applied the Delayed Rejection Adaptive Metropolis (Haario *et al.*, 2006; Smith, 2013), a Metropolis Hastings type Markov Chain Monte-Carlo (MCMC), algorithm to verify that our deterministic results are reasonable. MCMC is a widely used sampling method which allows the study of sample point distributions per the evaluation of iteratively generated random samples, each strictly dependent on its prior. More specifically, DRAM is an acceptance-rejection algorithm accepting, or rejecting, newly generated parameters during each iteration based on their ability to satisfy a higher likelihood than the current evaluated parameter. If a parameter is rejected, delayed rejection permits the further evaluation of other parameters (Smith, 2013). For this analysis we used a normal joint *a priori* distribution with the mean obtained from the point estimate obtained using the Levenberg-Marquardt gradient based optimization method (Kelley, 1999). The *a priori* estimates are used as initial values in the DRAM algorithm, in turn calculating a posterior distribution which allows us to study the potential of pairwise correlations and possible impacts on identifiability.

## Results

***Sensitivity, subset selection and comparison of model predictions:*** In this study we selected two subsets of parameters to analyze. The first is based on known parameters of physiological interest, whereas the second is formed to include identifiable parameters determined from local sensitivity and structured correlation analysis. Figure 3 presents the normalized ranked parameter sensitivities for a characteristic data set and the two parameter subsets chosen for this study for Patient 233. $R_{av}$ and $R_{mt}$ are not included in the subset as their ranked sensitivities are less than 0.01. These parameters were fixed at their nominal values and not included in correlation analysis. To justify fixing these parameters we doubled and halved them noting that changing their value has negligible effect on dynamic model predictions. For the remaining parameters, we used the structured correlation algorithm (Olufsen & Ottesen, 2013) to determine an identifiable parameter set. For all patients, we used a correlation threshold of $\gamma = 0.9$. This analysis determined the following sensitivity-based identifiable parameter subset including:

$$\hat{\theta} = \{\lambda_{rv}, \lambda_{lv}, E_{pa}, E_{pu}, R_{pul}, E_{sa}, R_{sys}, R_{pv}\}.$$

When this analysis was run on other patient data in the study the same identifiable subset emerged although in some cases the ranking order changed slightly. Results shown in Figure 4, for patient 233, are obtained by estimating the sensitivity-based identifiable subset $\hat{\theta}$ keeping all other parameters fixed at their nominal value. Results shown in Figure 4 are obtained by minimizing the least square error as presented in Eqn. (28). The optimized parameter values $\hat{\theta}_{opt}$ are given in Table 3.

Table 3. Nominal and optimized parameter values for patient 233 using the sensitivity-based identifiable parameter subset, $\hat{\theta}$ and the physiological-based parameter subset, $\tilde{\theta}$.

| Parameter | Units | Nominal | Optimized SBIP, $\hat{\theta}$ | Optimized PBP, $\tilde{\theta}$ |
|---|---|---|---|---|
| $\lambda_{lv}$ | 1/mL | 0.03 | 0.0289 | --- |
| $\lambda_{rv}$ | 1/mL | 0.025 | 0.0183 | --- |
| $E_{lv}$ | mmHg/mL | 3.51 | --- | 2.704 |
| $V_{d,lv}$ | mL | 10 | --- | 8.13 |
| $E_{rv}$ | mmHg/mL | 0.92 | --- | 3.65 |
| $E_{pa}$ | mmHg/mL | 0.39 | 0.1696 | 0.2074 |
| $E_{pu}$ | mmHg/mL | 0.30 | 0.212 | 0.1059 |
| $R_{pul}$ | mmHg · s/mL | 0.095 | 0.0481 | 0.05015 |
| $E_{sa}$ | mmHg/mL | 0.820 | 0.678 | 0.767 |
| $R_{sys}$ | mmHg · s/mL | 0.835 | 0.637 | 0.715 |
| $R_{pv}$ | mmHg · s/mL | 0.029 | 0.00797 | 0.00777 |

Results for patient 233 are shown in Figure 5 estimating the physiologically-based parameter subset $\tilde{\theta} = \{E_{rv}, E_{lv}, R_{sys}, E_{pu}, R_{pul}, E_{pa}, R_{pv}, E_{sa}, V_{d,lv}\}$. The optimized parameter values for this subset are given in Table 3. Note that the two subsets $\hat{\theta}$ and $\tilde{\theta}$ have 6 parameters in common, $\theta_{com} = \{R_{sys}, E_{pu}, R_{pul}, E_{pa}, R_{pv}, E_{sa}\}$. From a physiological point of view, the systolic elastances of the left and right ventricles ($E_{lv}$ and $E_{rv}$) are easier to interpret than the parameters defining diastolic filling ($\lambda_{lv}$ and $\lambda_{rv}$), and therefore are included in the physiological subset $\tilde{\theta}$. However, correlation analysis revealed that parameters $E_{rv}$ and $\lambda_{rv}$ in addition to $E_{lv}$ and $\lambda_{lv}$ are correlated and that the elastance parameters are less sensitive. Therefore, the identifiable parameter set $\hat{\theta}$ includes $\lambda_{lv}, \lambda_{rv}$ and fixes $E_{lv}$ and $E_{rv}$ at their nominal values. Next, we observe that $V_{d,lv}$ is strongly correlated with $E_{rv}, E_{lv}, R_{sys}, E_{pu}$ and $E_{sa}$. Removing $V_{d,lv}$ from the subset, fixing it at its nominal value, eliminated correlations within the remaining subset.

When we optimize $\hat{\theta}$ and $\tilde{\theta}$, the residual least squares cost are both small (e.g. $1.5 \times 10^{-5}$ and $1.67 \times 10^{-3}$ respectively for patient 233) but the residual is smaller with the sensitivity based identifiable parameter subset. In addition, the physiological-based parameter subset, $\tilde{\theta}$, for patient 233 gives estimations of left and right ventricular volume that differ significantly in magnitude (Figure 5f) while in the sensitivity-based identifiable parameter subset, $\hat{\theta}$, optimized model the systolic and diastolic volumes for the ventricles are similar in magnitude (Figure 4f). Systolic and diastolic volumes for the right and left ventricles are expected to be similar in a normal cardiovascular state (Alfakih *et al.*, 2003), which is in line with the sensitivity-based identifiable parameter subset optimization predictions; however, no studies have been performed quantifying relative ventricular volumes in post-transplant hearts.

***Parameter identifiability:*** We performed a delayed rejection adaptive metropolis (DRAM) analysis to confirm the degree of identifiability for each of our two parameter subsets and to further uncover which parameters in the subset are correlated. For each of the two parameter subsets $\hat{\theta}$ and $\tilde{\theta}$ we set up normal joint *a priori* distribution with the mean obtained from the point estimates discussed above. The DRAM algorithm was run with 100,000 sample points. To ensure that our solutions converge to steady state prior to calculating posterior distributions and correlations we removed, the burn-in period was set to 10,000 sample points. Figure 6 shows that for both parameter subsets, the chains have converged (top two panels). The bottom panels in Figure 6 show the posterior distributions for each parameter in both subsets. For the physiological subset $\tilde{\theta}$, we observed that the parameter $V_{d,lv}$, for the dead space volume in the left ventricle, has an identical distribution as the parameter $E_{lv}$, the elastance of the left ventricle.

This distribution suggests that the parameters are correlated with a single valued relation; this is equivalent to saying their Pearson correlation is +1. A Pearson's correlation of +1 is indicative of a perfect positive linear relationship (Everitt & Skrondal, 2010) between the parameters which is confirmed in Figure 7 depicting pairwise distributions. Therefore, fixing one of the two parameters may improve the DRAM results of the physiologically-based parameter subset since they are not mutually identifiable. Additionally, posterior and pairwise distributions for the sensitivity-based identifiable parameter subset confirm local observations that all parameters are independent.

***Longitudinal analysis:*** Using the cardiovascular systems model with the sensitivity-based identifiable parameter subset, RHC measures were analyzed to represent cardiovascular functional changes during recovery in five of the ten patients where multiple RHC measures were available. Figure 8 shows the changes in cardiovascular function, as indicated from simulated left and right ventricular pressure-volume loops and left and right ventricular power output, for two representative patients (patient 266 and 558). Ventricular power output can be calculated by integrating the area inside the pressure-volume loop and multiplying it by heart rate. The adoption of left ventricular power as a clinical measure is becoming more common (Cotter *et al.*, 2003b; Fincke *et al.*, 2004) and has been shown to be elevated at rest in septic shock and congestive heart failure with hypertension (Cotter *et al.*, 2003a). Figure 8 shows a consistent reduction in right and left ventricular pressure as well as left and right ventricular power output during recovery in patient 266 that are not seen with patient 558. For patient 266, the systolic volumes in both ventricles increases during recovery with diastolic volumes remaining relatively constant. However for patient 558, systolic volumes in both ventricles show a great degree of variability with a general reduction in volume during recovery and only the left ventricular diastolic volume remains nearly constant. This representation of cardiovascular function cannot be obtained from the RHC measures without the use of the computational analysis. As shown in Figure 9, we can also track individual parameters longitudinally; the values of the sensitivity-based identifiable parameter subset are plotted for patients 266 and 558. The remaining three patients with longitudinal RHCs during recovery (363, 456 and 572, not shown) showed small increases in left ventricular cardiac power during recovery of 6, 27 and 22% rising to 1.15, 1.42 and 1.73 W, respectively. All of these values are well below the peak left ventricular power of 2.5 W for patient 558 indicating left ventricular cardiac power below 2W during recovery is favorable.

## Discussion

In this study, we develop an analysis methodology where a series of RHC measurements from recovering heart transplant patients are analyzed with an identifiable model of cardiovascular system dynamics to represent cardiovascular function progression during recovery. In the first portion of this study we carefully assessed which model parameters can be reliably identified with the sparse clinical data from EHRs including RHC measurements, systemic arterial blood pressure, heart rate, and other biometrics from the patient. A sensitivity-based parameter subset was selected and compared to an alternate parameter subset to show differences in parameter identifiability and correlation. The sensitivity-based parameter subset was shown to be identifiable and composed of independent parameters in the neighborhood of parameter values optimized to fit clinical data for one exemplary RHC dataset. In the second portion of this study we used this model to analyze recovery progression for 5 patients that had a series of RHCs over the span of 7 to 392 days post-transplant. Trends in simulated pressure-volume loops, left and right ventricular power output and model parameters over the course of recovery show differences in recovery progression. Two patients (266 and 558) were selected from the analysis that show distinct differences in recovery.

***Longitudinal analysis of cardiovascular function***. It is known that pulmonary hypertension (>25-30 mmHg) is often observed post-transplant with a trend to normal right-side pressures (~20 mmHg) in a successful recovery (Delgado *et al.*, 2001). We have analyzed RHC datasets at several post-transplant time points in patients 266, 363, 456, 558 and 572 to see what other cardiovascular metrics may be useful to predict outcome. We focused on patient 266 to represent what appears to be a successful heart transplant recovery.

For patient 266, left and right PV loop trends as a function of recovery time represented in Figure 8 shows a reduction in right ventricular pressure along with an increase in end systolic volume over seven time points spanning an 11-month period post-transplant. The RHC data alone shows this reduction in ventricular pressures. However, we observe that over time, the ejection fraction decreases by 5%, the left ventricular end diastolic pressure decreases from about 10 to 6 mmHg, the systemic arterial elastance decreases (reduction in stiffness) by 54% and the systemic resistance decreases by 17%. Even more interesting is the trend in left and right ventricular power output. As recovery progresses, patient 266 experience a decrease in left ventricular power output from 1.8 W at day 57 post-transplant to 1.4 W at day 392. These predicted metrics describe the constellation of concurrent changes that can be quantified in the transplanted heart, but also accounts for changes in the pulmonary and systemic cardiovascular system over time. The trend in ventricular power can be interpreted as the heart is working less to maintain cardiac output as the patient recovers. A normal left ventricular power output at rest is around 1 W (Cotter *et al.*, 2003b; Klasnja *et al.*, 2013) and roughly represents a blood pressure of 120/80 mmHg, a heart rate of 70 beats/min and a stroke volume of 70 mL. While the predicted reduction in ejection fraction points to reduced cardiovascular function in this patient, this is likely due to the fact that volumes are not constrained by any clinical measures in this study. The inclusion of periodic echocardiograms along with the heart rate close to the time of each RHC could be used as another set of measures to bound the left ventricular end diastolic and end systolic volumes in the optimized models.

Patient 558, on the other hand, shows a slight increase in right ventricular pressure over the seven time points spanning 5 months. For this patient, the model predicts a consistent left ventricular end diastolic pressure of about 11 mmHg, a decrease in systemic arterial elastance of 28%, a large increase in systemic resistance of 80% as well distinct arterial hypertension rising to 180 mmHg as shown in the left ventricular pressure model output. Markedly, the left ventricular power does not show any downward trends and remains high over the observed recovery period varying between 1.5 and 2.5W. The predicted increase in systemic resistance for patient 558 is slightly below the range of 120% and 250% increase from normal healthy values of systemic resistance observed in patients with congestive heart failure with hypertension and pulmonary edema, respectively (Cotter *et al.*, 2003a) but still reflects a negative trend during recovery. It may also be noted that our computation of cardiac power from generated pressure volume loops takes into account the diastolic filling pressure, which is ignored in clinical calculations making our prediction of cardiac power a better estimate in the case of pulmonary hypertension and congestive heart failure when diastolic filling pressures increase. A comparison of cardiovascular metrics between patients 266 and 558 suggest a less successful recovery for patient 558.

To test if these retrospective predictions align with actual patient outcome, the EHRs for each of the longitudinally tracked patients were checked in March of 2019. This represented post-transplant time points of 54 and 49 months for patients 266 and 558, respectively. Patient 266 exhibited complications due to osteoarthritis, likely precipitated from long-term immunosuppression but had no cardiovascular related complications. Cardiovascular function checked 43 months post-transplant showed normal cardiovascular function with ejection fraction at 60% and a stroke volume of 54 mL. In contrast, patient 558 at the 48-month post-transplant time point exhibited biventricular failure with an ejection fraction at 49%, right atrial pressure at 25 mmHg, pulmonary hypertension with right ventricular pressures of 48/35 mmHg, systemic hypertension at 124/98 mmHg, a heart rate of 123 beats/min and diminished stroke volume of 23 mL. The remaining longitudinally tracked patients (363, 456 and 572) predicted to have a positive outcome had no cardiovascular complications noted in their EHRs at post-transplant time points of 47, 46 and 52 months, respectively. Even though the small number of patients tracked in this study does not yet validate this approach it does illustrate the utility of mechanistic computational analysis of clinical data as tool for clinicians to use as heart transplant patient recovery is assessed.

***Model parameter subset selection***. The model selected for this study was developed to estimate the systolic and diastolic pressures, average pulmonary capillary wedge pressure and cardiac output obtained from EHR records. Using the Smith et al. (2004) cardiovascular systems model as a reference, we carefully selected only the model components that could be informed by the available data. The major components omitted were ventricular-ventricular interaction and inertance at each of the heart valves. A more complex cardiovascular system model could be implemented per the availability of pressure time courses from the RHC measurements and/or the addition of echocardiography data measuring volumes in the left ventricle.

This study compared two parameter subsets. The first was selected based on knowledge of the cardiovascular system and quantities of interest, the second was selected using sensitivity analysis and subset selection. The majority of parameters were present in both subsets, yet rigorous analysis revealed that the physiological-based subset included correlated parameters. The physiologically-based parameter subset included resistances of the systemic vasculature ($R_{sys}$), pulmonary vasculature ($R_{pul}$) and pulmonary valve ($R_{pv}$) along with elastances of the left ventricle ($E_{lv}$), right ventricle ($E_{rv}$), pulmonary vein ($E_{pu}$), pulmonary artery ($E_{pa}$), systemic arteries ($E_{sa}$) and the dead space volume in the left ventricle ($V_{d,lv}$) as shown by the red squares in Figure 3. The sensitivity-based subset selection approach omitted $V_{d,lv}$, while identifying the correlation between the elastance parameters in the left and right ventricles and the more sensitive diastolic filling exponents in the left and right ventricle, $\lambda_{lv}$ and $\lambda_{rv}$ as shown by the blue circles in Figure 3. This result suggests that quantifying the parameters determining diastolic filling in the left and right ventricle will have more ability to discriminate underlying cardiovascular function than quantifying the parameters determining systolic contraction. The sensitivity-based approach did not identify the pulmonary valve resistance ($R_{pv}$) as having high sensitivity, however, this parameter was added to the sensitivity-based identifiable subset since we had data for right ventricular and pulmonary arterial pressure across the valve along with cardiac output. While it is possible to uniquely identify this parameter solely from data, it was included as an adjustable parameter to provide maximum flexibility in matching the two pressures and cardiac output simultaneously.

The physiologically-based approach for selecting model parameters to optimize relies on intuition to determine adjustable parameters of interest. However, this approach does not reveal correlations between model parameters, and therefore can lead to subsets which are not uniquely identifiable given the model and associated experimental data. The sensitivity-based approach selected a subset with eight parameters, six of which were also included in the physiologically-based parameter subset. In addition to parameter estimation, the sensitivity analysis and subset selection methods employed here can also be used for experimental design, e.g. to analyze what output quantities are needed to estimate specific parameters. The model studied here was simplified compared to the model by Smith et al. and a major factor ignored was VVI. This component is believed to be important for patients with severe pulmonary hypertension however with clinical measures only from the right ventricle used in this study incorporating VVI with the Smith et al. model components would likely lead to insensitive parameters in that component of the model.

Finally, it should be noted that in this study the two subsets were studied independently, i.e. the sensitivity-based analysis was purely informed by model analysis. Another approach is to pick parameters of interest and then test if the subset picked contains identifiable parameters. Finally, the local estimates can be validated by comparing the point-estimates determined using gradient-based method agree with the max of the distributions obtained using DRAM. For this study, this procedure was used for one dataset (Patient 233) and results agreed. Yet both the point-estimates and the parameter distributions were obtained subject to values of non-estimated parameters. In future studies, more work is needed to determine uncertainty to perturbation of fixed parameters within physiological bounds. This can be done using global sensitivity analysis combined with more simulations varying these and quantifying how variation in fixed parameters impact outcomes.

***Model optimization using both selected parameter subsets***. Model optimization using both the sensitivity-based and the physiologically-based parameter subsets were able to fit the RHC data and systemic arterial pressure measures, as can be seen by comparing the RHC measures (dashed lines) and simulated pressure and cardiac output time courses (solid lines) in panels a, b and c of Figures 4 and 5, respectively. The main difference between the two subsets are in their predictions of the left and right ventricular volume. The physiologically-based subset predicts smaller right ventricular volumes and larger left ventricular volumes than the sensitivity-based subset as seen in Figures 4e, f and 5e, f. The model predictions obtained using the sensitivity-based parameter subset leads to predicted left and right ventricular volume that are closely matched. It has been observed in normal hearts that the ventricular volumes are typically similar between right and left sides of the heart (Alfakih *et al.*, 2003; Hudsmith *et al.*, 2005; Hergan *et al.*, 2008), with the right ventricular volume on the order of 5-10% smaller than the left ventricular volume. However, in these same studies right ventricular volumes can be as much as 48% smaller or 68% larger depending on the gender considered, the measurement modality and the method used to calculate the volume from the images obtained. No similar studies have been conducted for ventricular volumes in patients with cardiac hypertrophy, pulmonary hypertension or after heart transplantation.

Confidence in the use of the sensitivity-based identifiable parameter subset is further bolstered by the results of the DRAM analysis as shown in Figure 6. In the neighborhood of the model optimization to patient 233 data, parameter distributions for all parameters in the selected sensitivity-based identifiable subset show a narrow Gaussian distribution whereas the parameter distributions for two parameters in the physiologically-based parameter subset, the pairwise distributions for $E_{lv}$ and $V_{d,lv}$, are clearly correlated and their distributions are proportional over the range of each parameter value.

***Optimized parameter relationship to CV function and subsequent model predictions***. Optimization of selected model parameters tell us about the underlying function that represents the upper level phenotype quantified clinically in this case with RHC and systemic arterial blood pressure measures. The optimized model parameters of the first complete RHC dataset from the ten patients in this study are shown in Table 4, we observe that the five patients (66, 233, 558, 572 and 839) with values for $\lambda_{lv}$ below 0.03 have the greatest left ventricle diastolic filling. Comparing these patients with the remaining five patients we see mean left ventricular end diastolic volumes of 178 ± 23 mL versus 141 ± 23 mL indicating a strong relationship between $\lambda_{lv}$ and left ventricular end diastolic volume.

Table 4. Optimized parameter values for first post-transplant RHC dataset from each patient using the sensitivity-based identifiable parameter subset ($\hat{\theta}$)

| Parameter | Units | Patient Number | | | | | | | | | |
|---|---|---|---|---|---|---|---|---|---|---|---|
| | | 60 | 66 | 233 | 266 | 363 | 456 | 558 | 572 | 794 | 839 |
| $\lambda_{lv}$ | $mL^{-1}$ | 0.0353 | 0.02046 | 0.0289 | 0.0328 | 0.0336 | 0.0329 | 0.0277 | 0.0290 | 0.0448 | 0.0283 |
| $\lambda_{rv}$ | $mL^{-1}$ | 0.0185 | 0.00882 | 0.0183 | 0.0303 | 0.0289 | 0.0217 | 0.017 | 0.0223 | 0.0316 | 0.0115 |
| $E_{pa}$ | mmHg/mL | 0.427 | 0.176 | 0.1696 | 0.463 | 0.243 | 0.6064 | 0.2405 | 0.4603 | 0.349 | 0.194 |
| $E_{pu}$ | mmHg/mL | 0.0811 | 0.0215 | 0.212 | 0.0581 | 0.157 | 0.0748 | 0.0562 | 0.0492 | 0.40062 | 0.0289 |
| $R_{pul}$ | mmHg·s/mL | 0.0471 | 0.08084 | 0.0481 | 0.113 | 0.147 | 0.0915 | 0.0558 | 0.0653 | 0.0472 | 0.0538 |
| $E_{sa}$ | mmHg/mL | 0.719 | 0.7105 | 0.678 | 1.4007 | 0.838 | 0.757 | 1.018 | 0.995 | 0.625 | 0.9546 |
| $R_{sys}$ | mmHg·s/mL | 0.975 | 0.8069 | 0.637 | 1.035 | 1.047 | 0.878 | 0.591 | 0.788 | 0.6034 | 1.058 |
| $R_{pv}$ | mmHg·s/mL | 0.00435 | 0.00428 | 0.00797 | 0.01085 | 0.0064 | 0.026 | 0.00361 | 0.00562 | 0.00657 | 0.01602 |

Since this model is nonlinear, it does not encode a one to one relationship between each model parameter and the upper level clinical measures. However, the power of this approach is that with a model tailored to describe an individual patient's cardiovascular system function, we can make predictions of cardiovascular system function that are not able to be easily measured. One example is the left and right ventricular pressure volume loops, which are often used to more accurately determine functional metrics when available in the clinic. These metrics include the left ventricular end diastolic and end systolic pressure volume relationships and both right and left ventricular work and power, which show important changes from health to dysfunction. Even though these are important diagnostic metrics, simultaneous measurement of left ventricular volume and pressure without using an invasive indwelling catheter in the left heart is impossible in the clinic, therefore they are rarely obtained. Employing our model analysis with the minimally invasive RHC measurements enables us to predict both right and left ventricular volumes while fitting the cardiovascular pressures in a patient-specific manner, resulting in these PV loops. In Figures 4d, e and f we see the predicted left and right ventricular pressure, volumes and pressure volume loops for patient 233. The relatively high left ventricular diastolic filling pressure shown, resulting from the pulmonary hypertension of the right side along with arterial systemic hypertension warrants close monitoring however the left ventricular ejection fraction of about 71% indicates an efficiently functioning heart.

## Conclusion

In this study, we have presented a workflow pushing model design and sensitivity analysis prior to model optimization in order to aid in selecting the model parameters that are most likely to be informed by the clinical measures. We have used clinical RHC and systemic arterial blood pressure data measured from patients, shortly after heart transplantation, to determine an identifiable parameter subset, and contrasted it to a subset that was selected based on physiological features of interest. Optimization on both parameter subsets replicated the clinical data equivalently. However, a single insensitive parameter in the physiologically-based subset decreased the confidence in the identification of the remaining parameters and the predictions made by the physiologically-based subset optimized model. To illustrate the potential to track cardiovascular function over time, model optimizations were performed on multiple RHC and systemic arterial pressure measurements for several patients and the trends in model parameters were shown. Predictions made by looking at model results based on this retrospective longitudinal data were confirmed by following up on these patients at the 4-year post-transplant time point illustrating the utility of this patient-specific model analysis. This approach has the ability to provide clinicians with previously unobtainable functional information, such as left and right ventricular pressure volume loops and systemic vascular resistance, from routinely obtained RHC measures. This additional functional information is not only valuable in the assessment of post-heart transplant recovery, but in other cases of cardiovascular dysfunction where RHC measurements are made such as heart failure both with reduced and preserved ejection fraction or pulmonary hypertension.

# Appendix A

*Model Equations*

The complete list of differential equations representing the rate of change of volume of the compartments in this study are as follows:

$$\frac{dV_{lv}}{dt} = Q_{mt} - Q_{av}$$

$$\frac{dV_{sa}}{dt} = Q_{av} - \frac{E_{sa}V_{sa} - E_{sv}V_{sv}}{R_{sys}}$$

$$\frac{dV_{vc}}{dt} = \frac{E_{sa}V_{sa} - E_{sv}V_{sv}}{R_{sys}} - Q_{tc}$$

$$\frac{dV_{rv}}{dt} = Q_{tc} - Q_{pv}$$

$$\frac{dV_{pa}}{dt} = Q_{pv} - \frac{E_{pa}V_{pa} - E_{pu}V_{pu}}{R_{pul}}$$

$$\frac{dV_{pu}}{dt} = \frac{E_{pa}V_{pa} - E_{pu}V_{pu}}{R_{pul}} - Q_{mt}$$

where

$$Q_{mt} = \begin{cases} \frac{\left(E_{pu}V_{pu} + P_{th}\right) - P_{lv}}{R_{mt}} & \text{if valve open } \left(P_{pu} > P_{lv}\right) \\ 0 & \text{otherwise (valve closed)} \end{cases}$$

$$Q_{av} = \begin{cases} \frac{P_{lv} - E_{sa}V_{sa}}{R_{av}} & \text{if valve open } (P_{lv} > P_{sa}) \\ 0 & \text{otherwise (valve closed)} \end{cases}$$

$$Q_{tc} = \begin{cases} \frac{E_{sv}V_{sv} - P_{rv}}{R_{tc}} & \text{if valve open } (P_{sv} > P_{rv}) \\ 0 & \text{otherwise (valve closed)} \end{cases}$$

$$Q_{pv} = \begin{cases} \frac{P_{rv} - \left(E_{pa}V_{pa} + P_{th}\right)}{R_{pv}} & \text{if valve open } \left(P_{rv} > P_{pa}\right) \\ 0 & \text{otherwise (valve closed)} \end{cases}$$

and

$$P_{lv} = A(\tilde{t})\left[E_{lv}\left(V_{lv} - V_{d,lv}\right)\right] + \left\{[1 - A(\tilde{t})]P_{0,lv}\left[e^{\lambda_{lv}(V_{lv} - V_{0.lv})} - 1\right]\right\} + P_{th}$$

$$P_{rv} = A(\tilde{t})\left[E_{rv}\left(V_{rv} - V_{d,rv}\right)\right] + \left\{[1 - A(\tilde{t})]P_{0,rv}\left[e^{\lambda_{rv}(V_{rv} - V_{0.rv})} - 1\right]\right\} + P_{th}$$

$$A(\tilde{t}) = e^{-a(\tilde{t} - T/2)^2}$$

# Figures

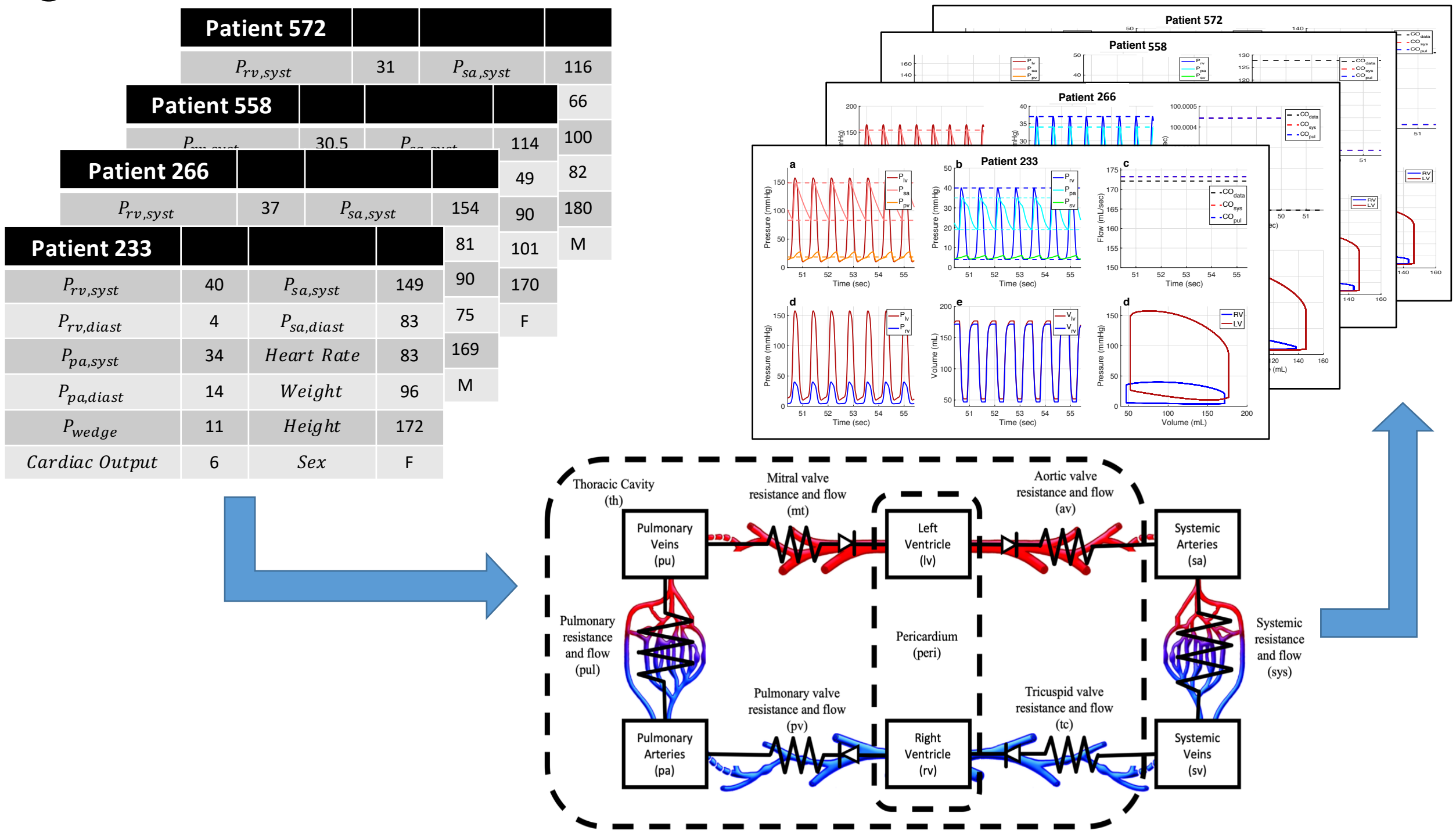

Figure 1. Retrospective approach where EHR data is used to identify a mechanistic model of the closed loop cardiovascular system thus generating model versions representing cardiovascular function of each patient (or patient at different times). These patient-specific instantiations of the model can then be used to understand differences between the outcomes across populations of patients.

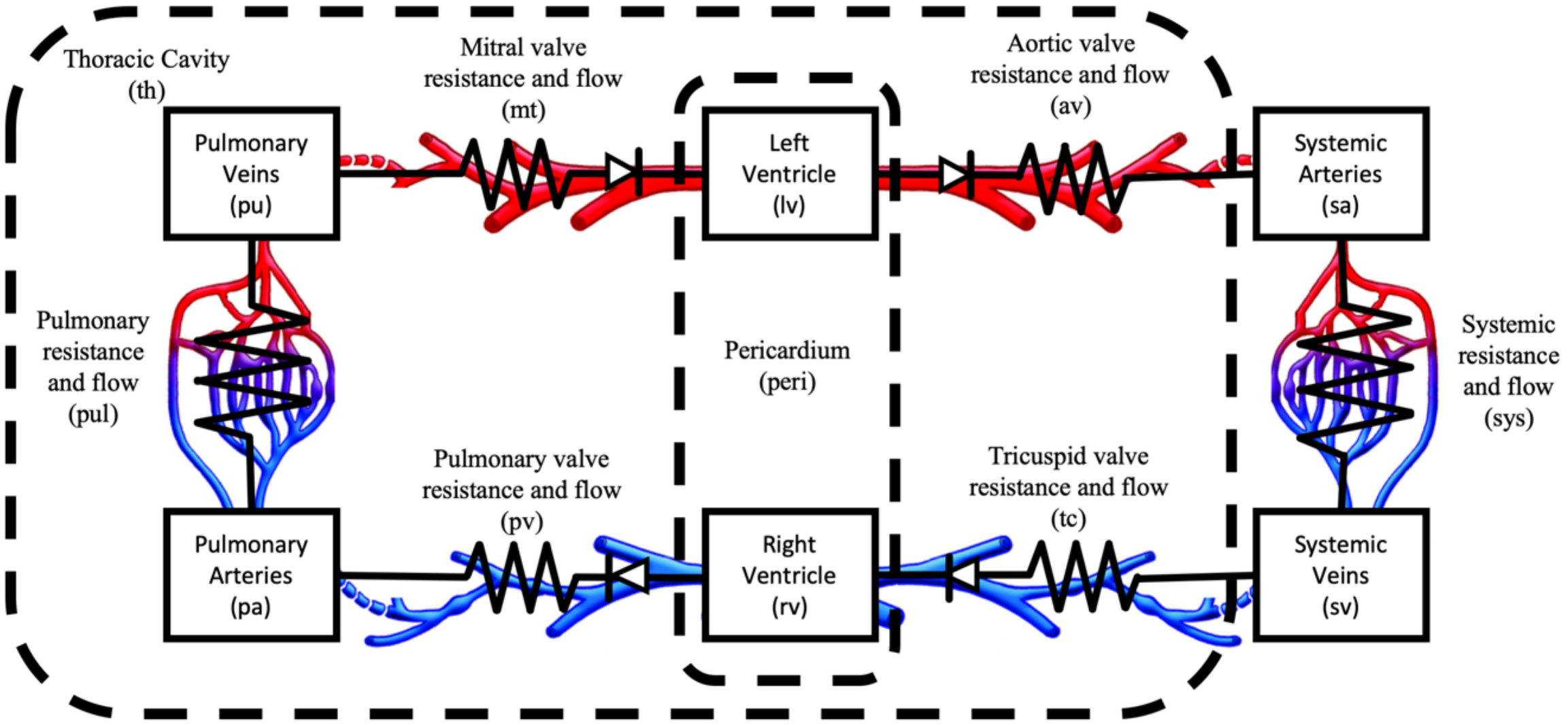

Figure 2. Schematic of closed loop cardiovascular system models used in this study. In the model by Smith et al. left and right ventricles interact by accounting for the dynamic pressure difference across the septal wall. In the reduced version of the model used in this study, ventricular-ventricular interaction along with inertance of the blood moving through the four heart valves is omitted.

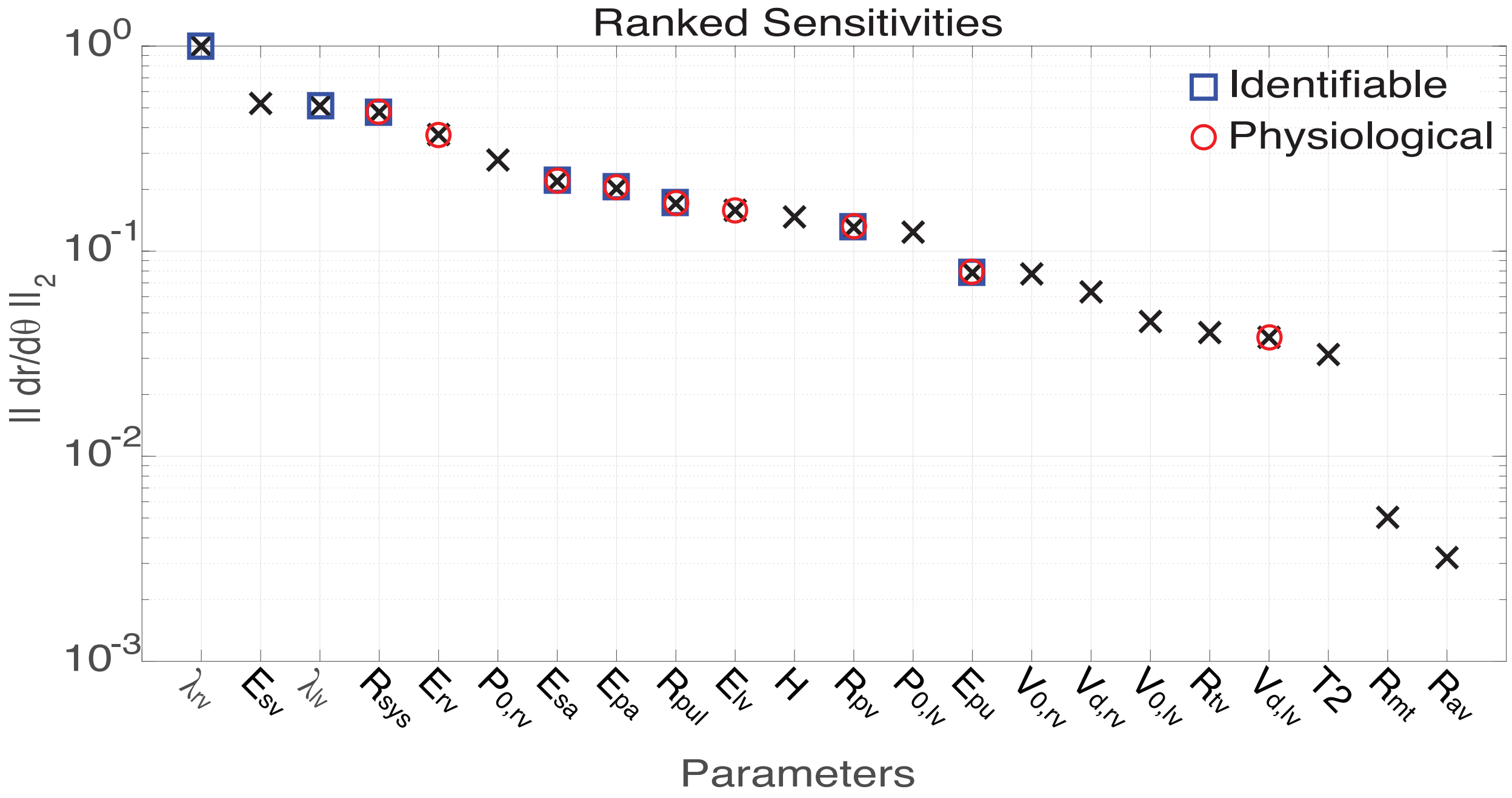

Figure 3. Ranked sensitivities for all parameters in the reduced model around nominal values for patient 233. Blue squares indicate the parameters selected by identifiability after sensitive parameters were also assessed for structural correlation while the red circles indicate those thought to be interesting based on their physiological meaning. Rank sensitivities for other patients in this study produces very similar rankings and yielded the same identifiable parameter subset.

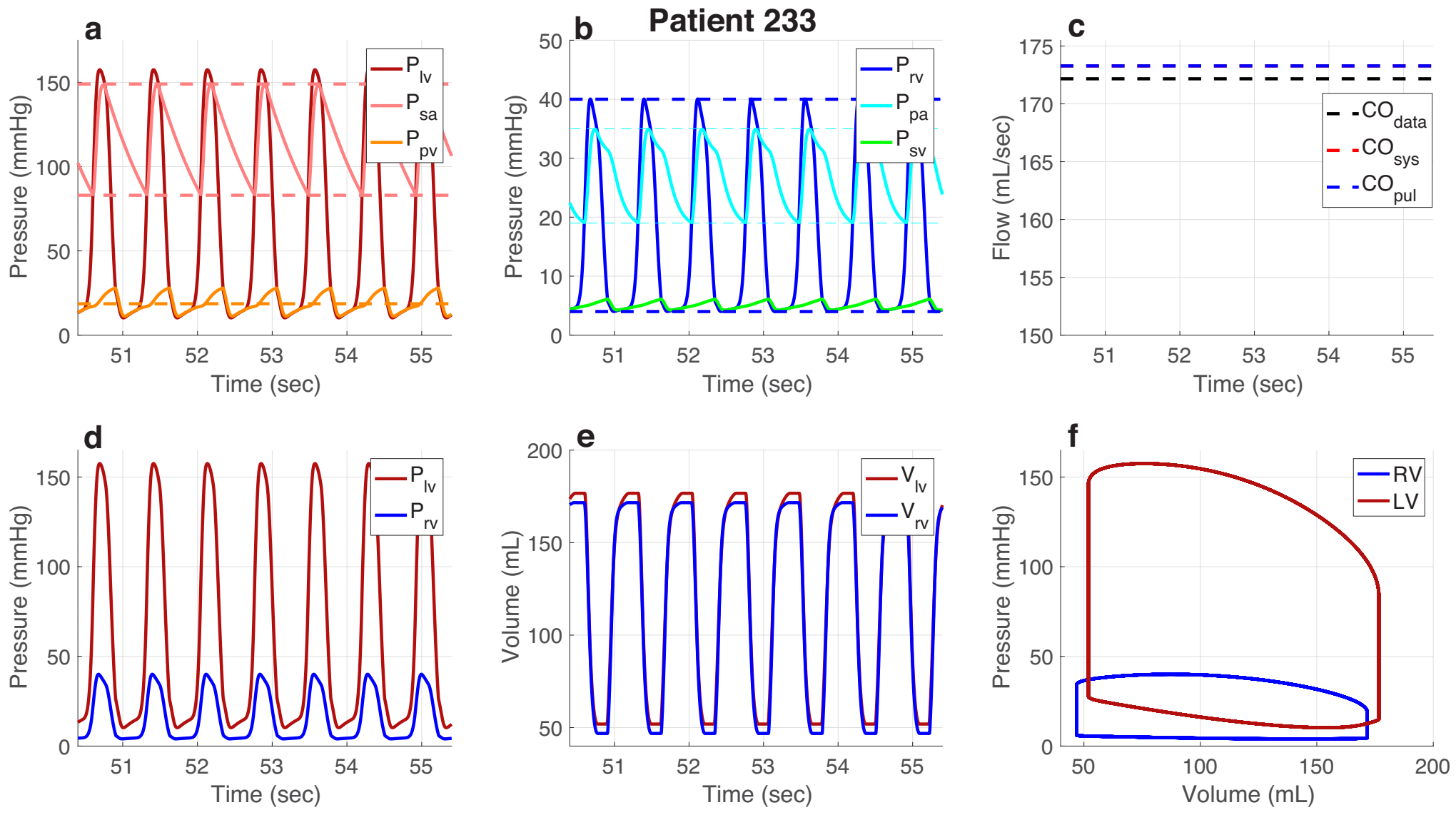

Figure 4. Model predictions for subject 233 with optimized sensitivity-based identifiable parameter subset values, $\hat{\theta}$. Left ventricle, pulmonary vein, & aorta (**a**) and right ventricle, pulmonary artery, & vena cava (**b**) comparison of computed results (solid lines) and data (broken lines). (**c**) comparison of computed cardiac output and data. **d**) comparison of left and right ventricular pressure. **e**) computed left and right ventricular volume (no data available). **f**) Left and right ventricular pressure volume loop (PV loop) with calculated stroke work.

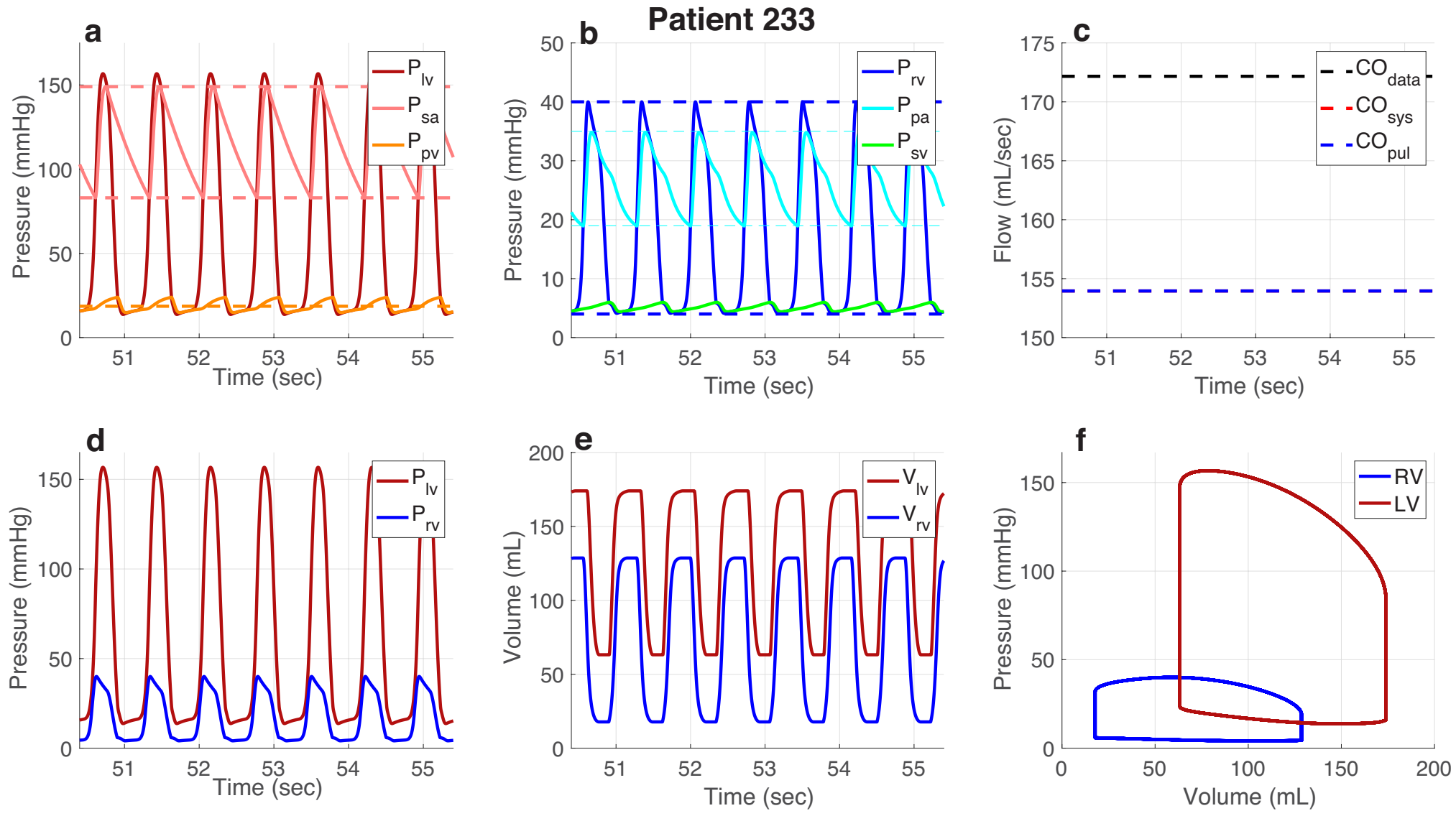

Figure 5. Model predictions for subject 233 with optimized physiologically-based parameter subset values, $\tilde{\theta}$. Left ventricle, pulmonary vein, & aorta (**a**) and right ventricle, pulmonary artery, & vena cava (**b**) comparison of computed results (solid lines) and data (broken lines). (**c**) comparison of computed cardiac output and data. (**d**) comparison of left and right ventricular pressure. **e**) computed left and right ventricular volume (no data available). **f**) Left and right ventricular pressure volume loop (PV loop) with calculated stroke work

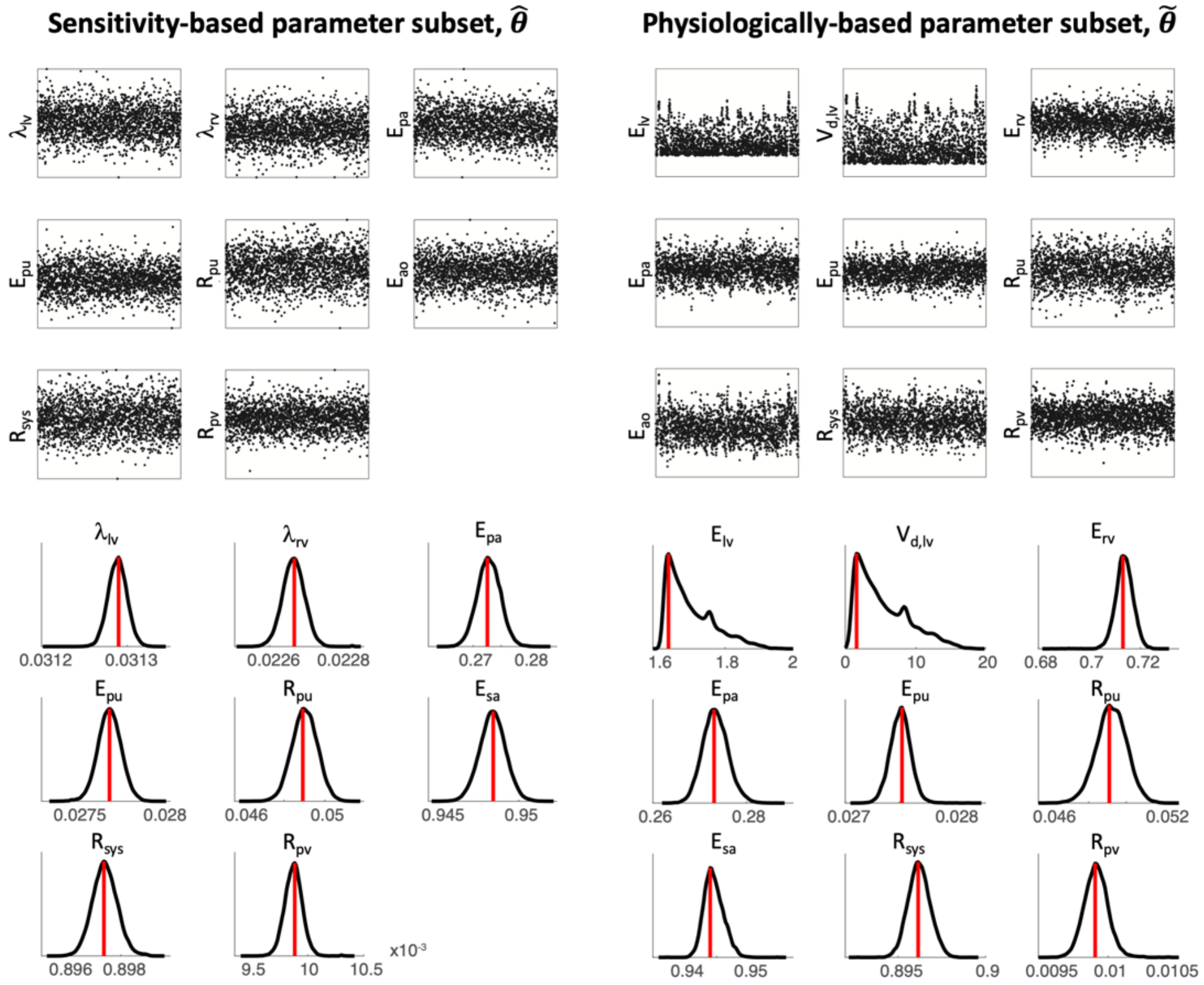

Figure 6. DRAM-based convergence chains (upper two panels) and parameter distributions (lower two panels) at optimized parameter values for patient 233 using the sensitivity-based identifiable (left column panels) and physiologically-based (right column panels) parameter subsets. Note that the parameter distributions for the sensitivity-based parameter subset are narrower than seen in the physiologically-based subset for all common parameters.

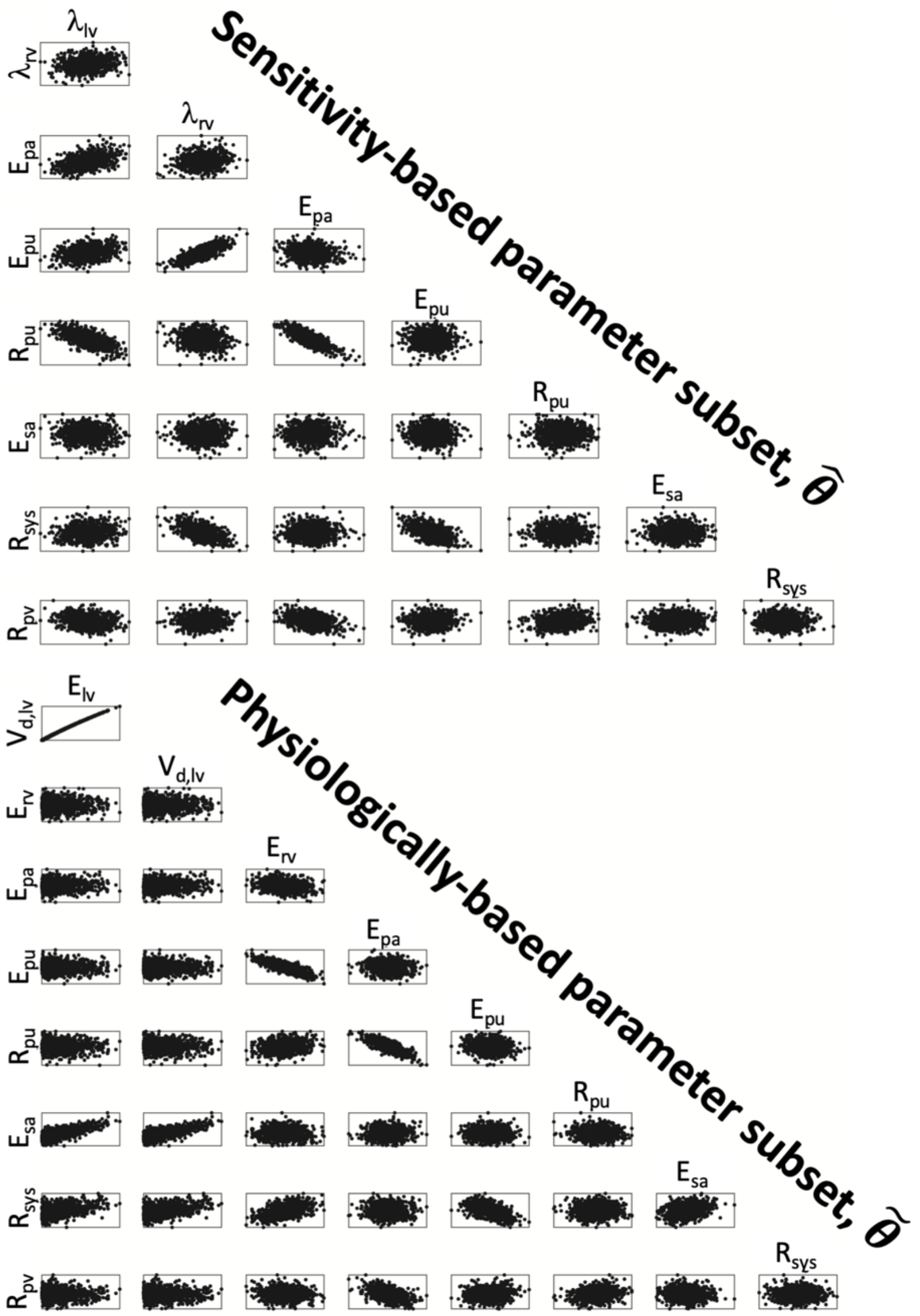

Figure 7. Pairwise distribution of parameters for each parameter subset to check for correlation. In the physiologically-based parameter subset, $E_{lv}$ and $V_{lv,d}$ are seen to be correlated since the relationship between the two distributions can be clearly seen. Pairwise parameters showing no correlation form a large cloud of points indicating no distinct relationship exists between the two distributions.

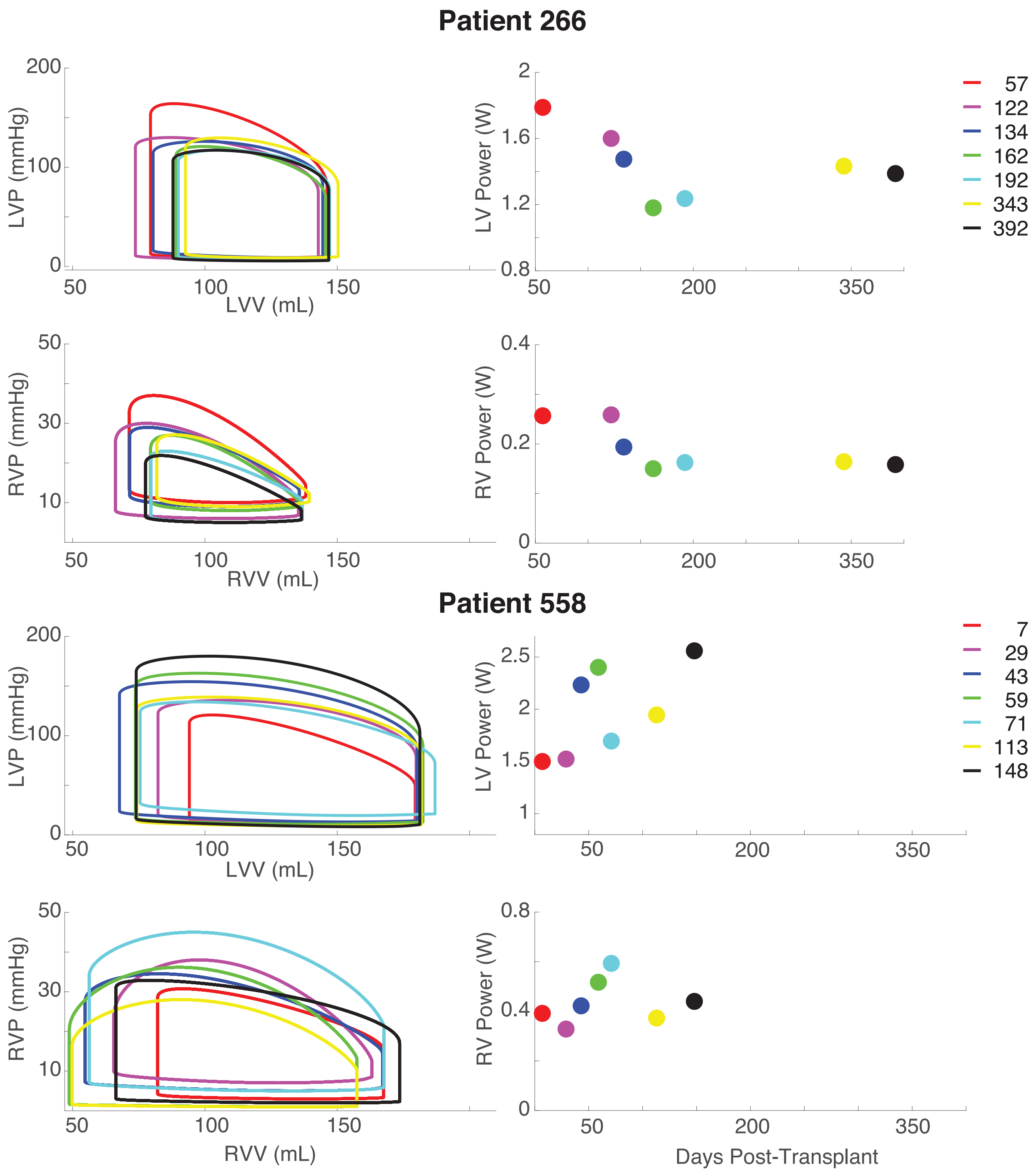

Figure 8. Simulated pressure-volume loops and left and right ventricular power output over time for patients 266 and 558 predicting dramatically different progression during recovery. Chronological progression earliest to latest by color is magenta, red, yellow, green, aqua, blue and black. Even if the ERH contains echocardiography and RHC data, the pressures from RHC and volumes from echocardiography are not obtained simultaneously therefore these pressure volume loops cannot be generated except through simulation as proposed here.

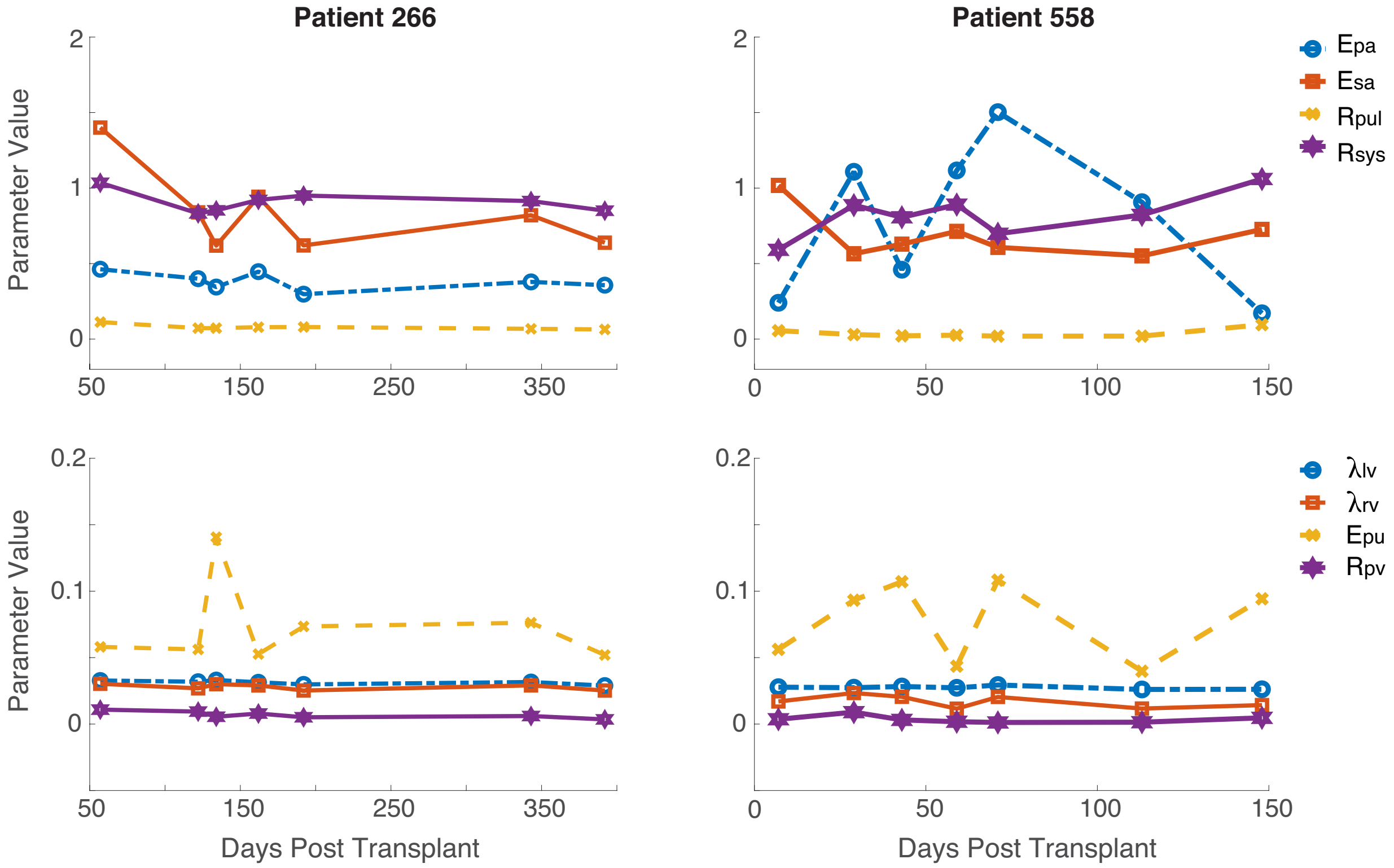

Figure 9. Longitudinal parameter trends for optimized parameters from the sensitivity-based identifiable parameter subset for patients 266 and 558.

# Supplemental Materials:
# Deep phenotyping of cardiac function in heart transplant patients using cardiovascular systems models

Amanda L. Colunga*, N. Payton Woodall*, Karam G. Kim*, Todd F. Dardas,
John H. Gennari, Mette S. Olufsen, Brian E. Carlson

## 1 Introduction

This document serves as a supplement for our manuscript "Deep phenotyping of cardiac function in heart transplant patients using cardiovascular systems models." A copy of the manuscript is available on ARXIV under the following identifier arXiv:1812.11857. A version of the model implemented in MATLAB (The MathWorks, Inc., Natick, MA) is available at github.com/alcolunga/Heart_Tx_CVS_Model and at https://wp.math.ncsu.edu/cdg/publications/. In this document we explore parameter convergence, parameter confidence intervals, and outlying data/parameter values.

## 2 Convergence

Once the established subsets of parameters have been optimized, we ensure that they are identifiable through convergence. Parameter estimation is repeated starting with eight initial parameter sets in which nominal parameter values are varied by 10%. Figure S1 shows convergence properties for each of our parameters in our subset $\hat{\theta} = \{\lambda_{lv}, \lambda_{rv}, E_{pa}, E_{pu}, R_{pul}, E_{sa}, R_{sys}, R_{pv}\}$.

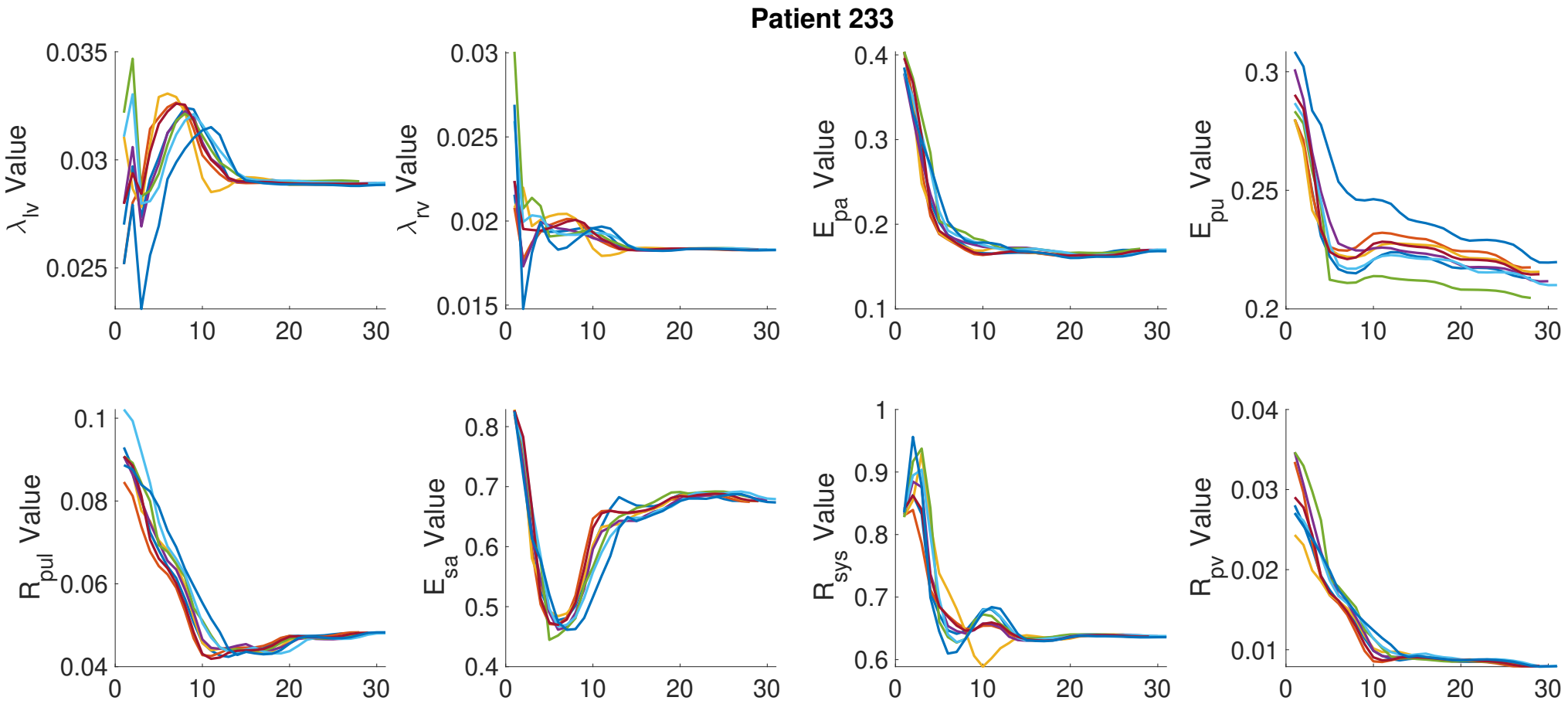

Figure S1: The convergence properties of each parameter in the subset $\hat{\theta}$ for Patient 233. The nominal parameters were initially varied by 10% for eight iterations and optimized. Note that parameter $E_{pu}$ has not optimized to one single value but only varies by $\sim 0.01$

Note that all parameters converge to a single value except the parameter $E_{pu}$ which has values varying between 0.21 and 0.22. The parameter sensitivity ranking in Figure S3 shows that $E_{pu}$ is the least sensitive parameter of $\hat{\theta}$. Coupled with the variability of $E_{pu}$ shown above, we determine that this parameter is non-influential which may lead to identifiability issues.

$E_{pu}$ is correlated to the computation of cardiac output, CO. The simulation was rerun adding a larger emphasis on the weight of the cost for CO. The heavier weight on the residual of CO allowed $E_{pu}$ to converge.

The convergence properties of $\tilde{\theta} = \{E_{es,lv}, V_{d,lv}, E_{es,rv}, E_{pa}, E_{pu}, R_{pul}, E_{sa}, R_{sys}, R_{pv}\}$ was also studied. Based on the sensitivity ranking $E_{es,lv}$ is quite sensitive and has wide variability of convergence seen in Figure S2. This may be an indicator that $E_{es,lv}$ is unidentifiable.

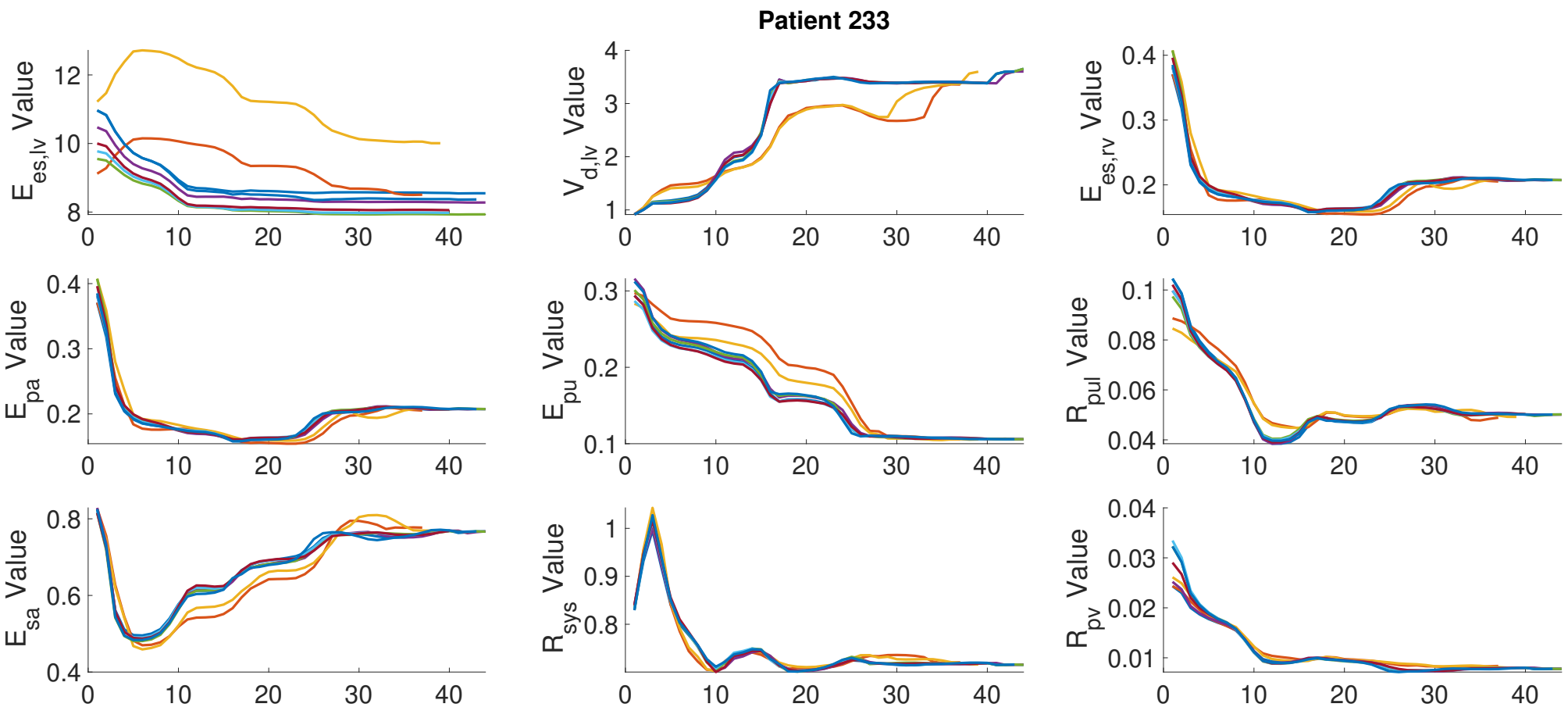

Figure S2: The convergence properties of each parameter in the subset $\tilde{\theta}$ for Patient 233. The nominal parameters were initially varied by 10% for eight iterations and optimized. Note that parameter $E_{es,lv}$ does not converge and varies widely between 8 and 11.

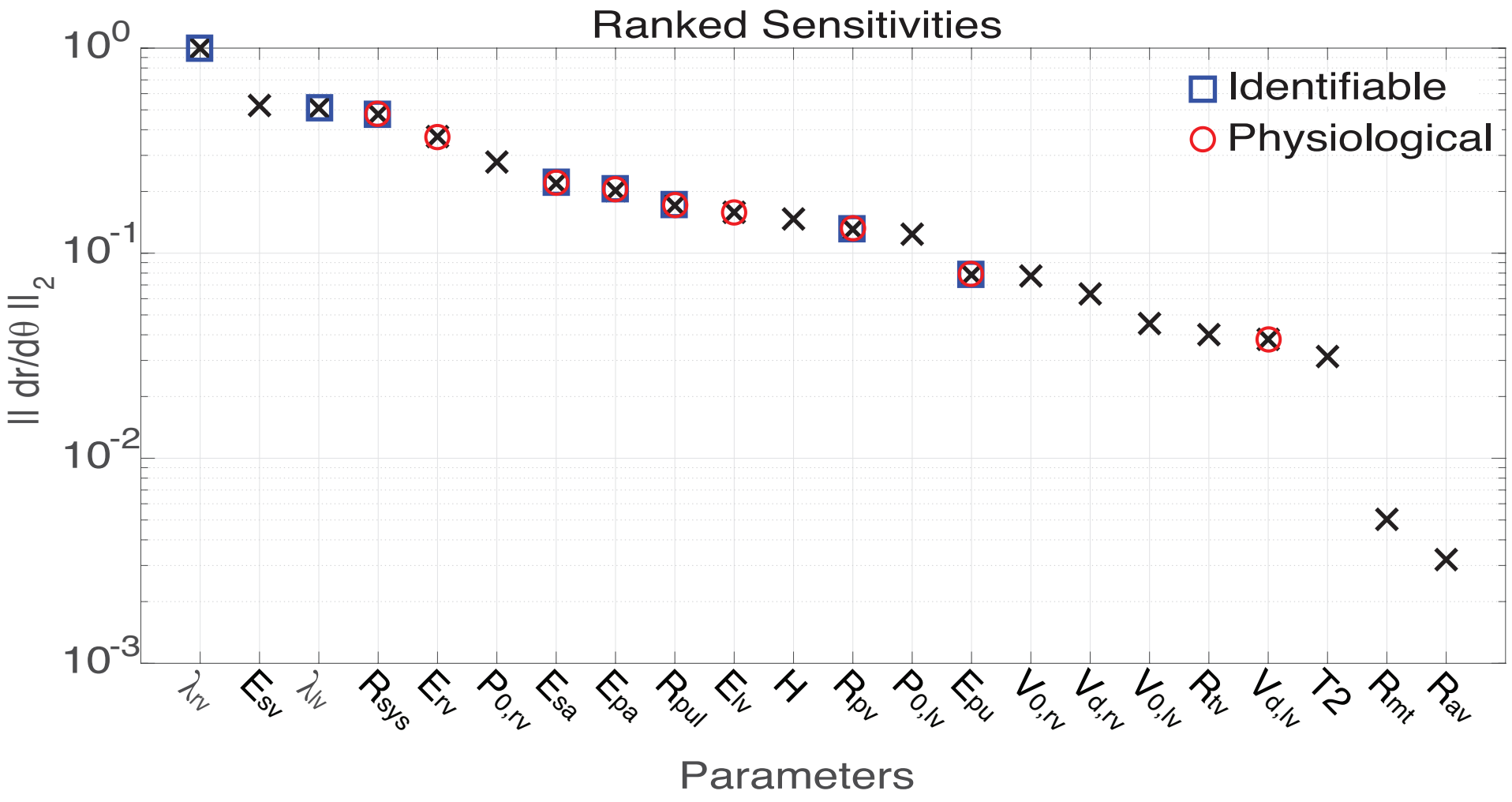

Figure S3: Ranked sensitivities for all parameters in the reduced model around nominal values for patient 233. Blue squares indicate the parameters selected by identifiability after sensitive parameters were also assessed for structural correlation while the red circles indicate those thought to be interesting based on their physiological meaning. Rank sensitivities for other patients in this study produces very similar rankings and yielded the same identifiable parameter subset.

## 3 Parameter Confidence Intervals

Parameter confidence intervals provide insight to the possible true parameter values of the computed optimized parameters. More specifically, a 95% parameter confidence interval it is not the probability, however, the confidence that in 95% of repeated experiments the parameters lie in the range. For the purpose of this experiment we were interested in computing the 97.5% parameter confidence intervals. Begin with the sensitivity matrix for the optimized subset $\hat{\theta}$ defined as

$$S_{i,j} = \frac{\partial y(t_i, \theta)}{\partial \theta_j} \frac{\theta_j}{y_i^d},$$

where $y(t_i, \theta)$ represents the model output at time $t_i$, $\theta_j$ denotes the j'th parameter of the $\hat{\theta}$ subset, and $y_i^d$ denotes the data measured at time $t_i$. Next, compute the inverse of the Fisher information matrix, $S^T S$, and find the diagonal elements.

$$D = diag\Big((S^T S)^{-1}\Big)$$

Following this, find the model residuals $R$ to compute the variance

$$\sigma^2 = \frac{1}{n-p} R^T R,$$

where $n$ is the number of time points $t$ and $p$ is the number of parameters in the subset. Now, compute the confidence intervals as

$$\text{confidence interval} = \hat{\theta} \pm \eta\delta$$

where $\eta$ is the Student's t-distribution value for 97.5% and $\delta = \sigma D$. Table S1 shows the 97.5% parameter confidence intervals for subset $\hat{\theta}$.

Table S1: 97.5% Parameter Confidence Intervals for Patient 233

| **Parameters** ($\times 10^{-2}$) | Optimized Value | 97.5% Confidence Interval |
|---|---|---|
| $\lambda_{lv}$ | 2.89 | [2.12 , 3.66] |
| $\lambda_{rv}$ | 1.83 | [1.62 , 2.04] |
| $E_{es,pa}$ | 17 | [16.10 , 17.81] |
| $E_{es,pu}$ | 21.2 | [16.90 , 25.51] |
| $R_{pul}$ | 4.81 | [04.18 , 5.44] |
| $E_{es,sa}$ | 67.8 | [67.05 , 68.53] |
| $R_{sys}$ | 63.7 | [63.38 , 64.07] |
| $R_{pv}$ | .797 | [0.13 , 1.46] |

## 4 Outlying Data/Parameter Values

The first RHC data for each patient were compared along with their respective computed nominal and optimized $\hat{\theta}$ parameters. Any patient with outlying values in two or more of the subsections were marked for analysis. Table S2 shows the means and standard deviations for the optimized $\hat{\theta}$ parameters. Visually, we determined outlying values by the use of box plots. Mathematically, an outlier is any value that is $>1.5$ times away from quartiles one or three. In summary, the equations are written in the form

$$Q1 = \nu_-,$$

$$Q3 = \nu_+$$

$$IQR = Q3 - Q1,$$

$$\eta_+ > Q3 + 1.5IQR$$

$$\eta_- < Q1 - 0.5IQR,$$

where $Q1$ is quartile one, $Q3$ is quartile 3, $\nu_{-/+}$ is the median of the lower or upper half of the data respectively, $IQR$ is the interquartile range and $\eta$ are the outliers.

Table S2: Parameter mean and standard deviation for the first RHC optimized parameters of every patient for subset $\hat{\theta}$.

| **Parameters** $(\times 10^{-2})$ | **Mean** $\pm$ **StDev** | **Parameters** $(\times 10^{-2})$ | **Mean** $\pm$ **StDev** |
|---|---|---|---|
| $\lambda_{lv}$ | $3.17 \pm .63$ | $R_{pul}$ | $7.73 \pm 3.35$ |
| $\lambda_{rv}$ | $2.19 \pm .77$ | $E_{sa}$ | $86.02 \pm 23.22$ |
| $E_{pa}$ | $34.83 \pm 15.07$ | $R_{sys}$ | $81.79 \pm 18.59$ |
| $E_{pu}$ | $12.34 \pm 11.68$ | $R_{pv}$ | $.841 \pm .699$ |

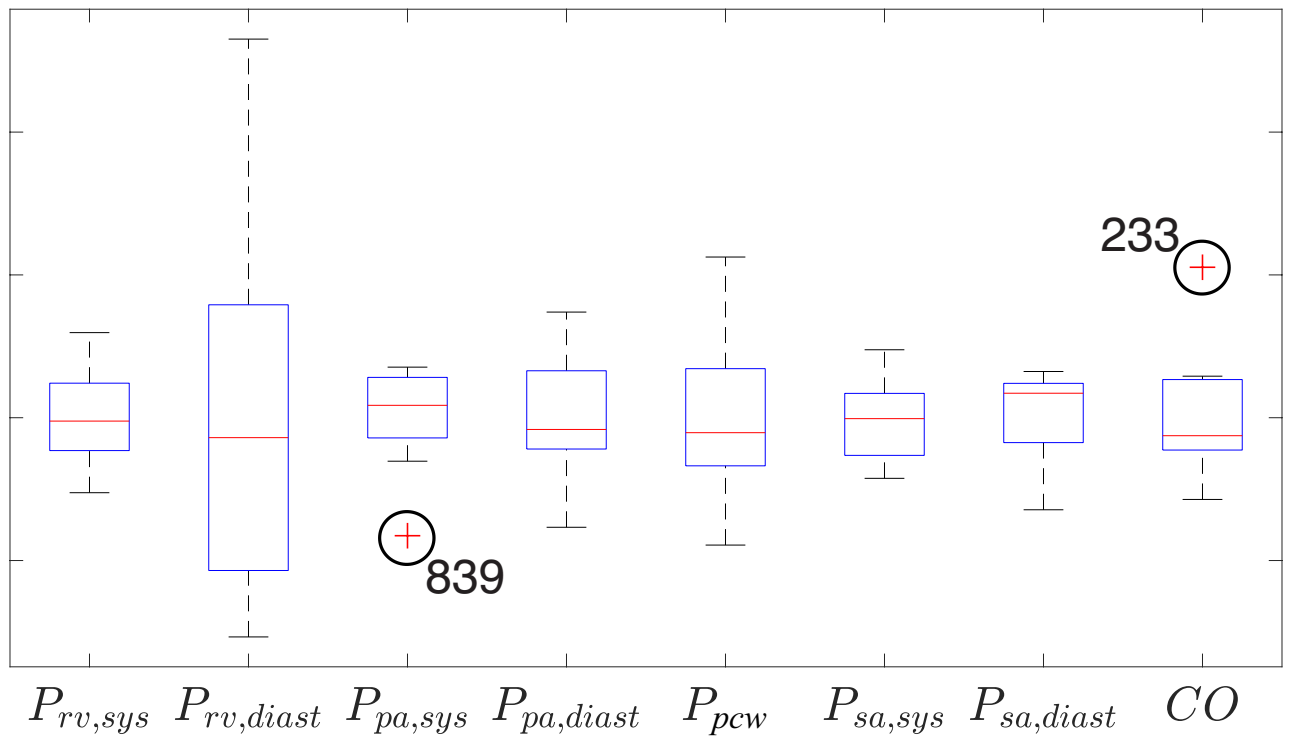

Figure S4: Box and whiskers plot of first RHC data of every patient. The plot is shown with the data normalized by the mean of each data point since they are on different orders of magnitude. Clear outlying data are marked as corresponding to their respective patient.

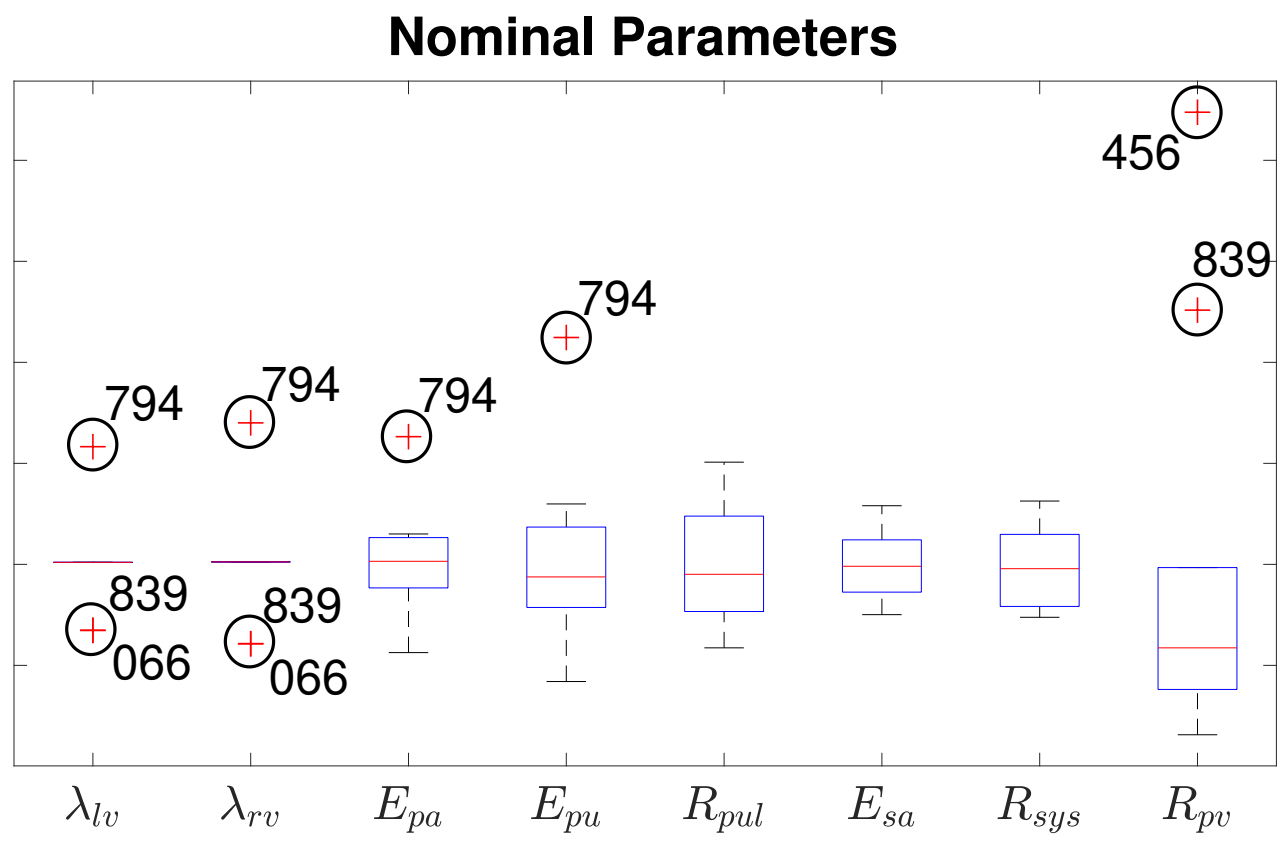

Figure S5: Box and whiskers plot of the nominal parameters of subset $\hat{\theta}$ for the first RHC data of every patient. The plot is shown with the parameters normalized by the mean of each parameter since parameters are on different orders of magnitude. Clear outlying parameters are marked as corresponding to their respective patient.

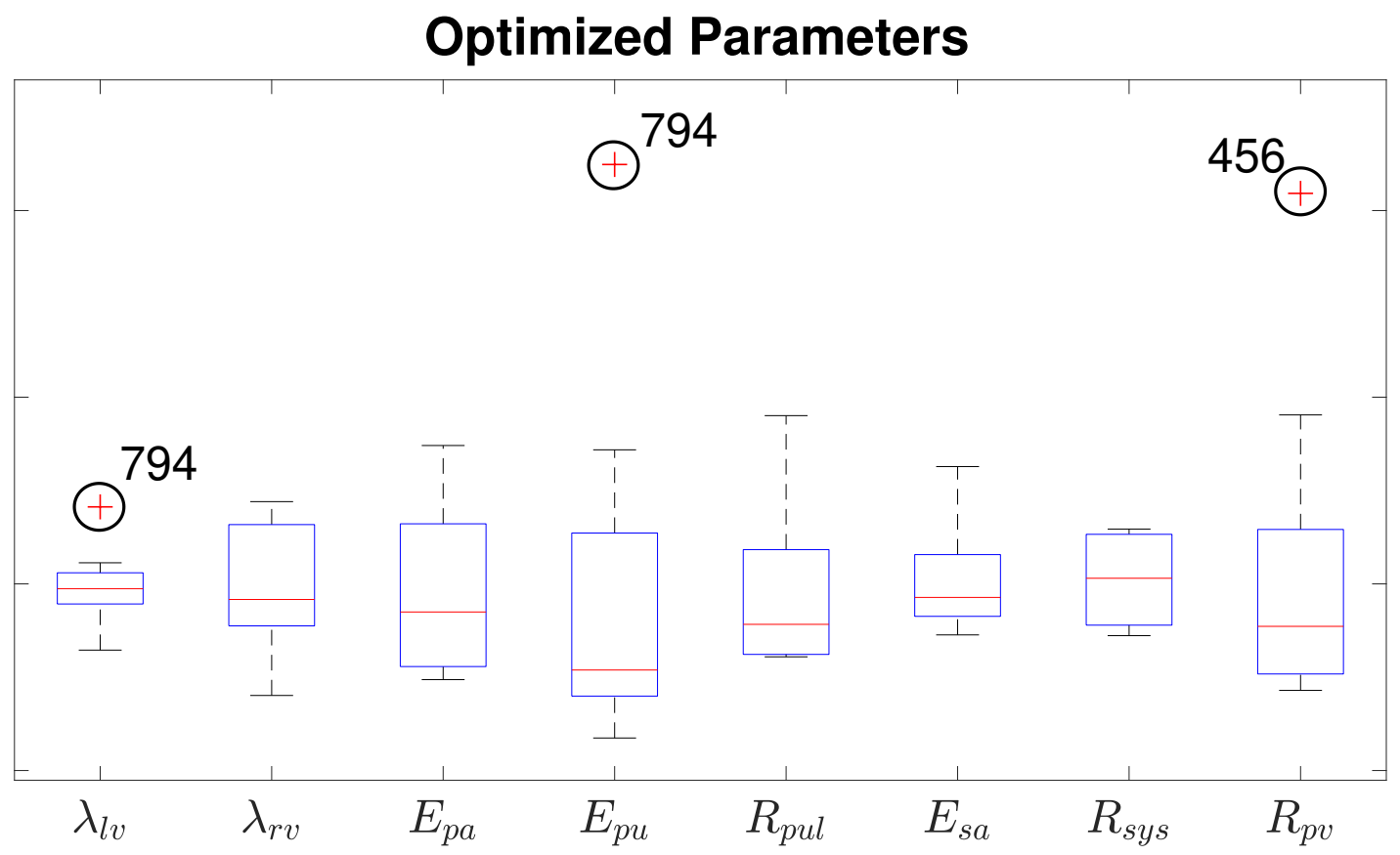

Figure S6: Box and whiskers plot of the optimized parameters of subset $\hat{\theta}$ for the first RHC data of every patient. The plot is shown with the parameters normalized by the mean of each parameter since parameters are on different orders of magnitude. Clear outlying parameters are marked as corresponding to their respective patient.

## 4.1 Analysis

Since parameters $\lambda_{lv}$ and $\lambda_{rv}$ were nominally set manually in order to ensure that our initial model matched the data, we will neglect any outliers for those parameters when looking at the nominal parameters box plot in Figure S5. Figures S4 and S5 show that patient 839 has an outliers in the data and nominal parameters. We raised the pulmonary arterial pressure at systole from 18 mmHg to 24.5 mmHg and this allowed the nominal pulmonary valve resistance parameter to lower from .0689 $\frac{\text{mmHg}\cdot\text{s}}{\text{mL}}$ to 0.0049 $\frac{\text{mmHg}\cdot\text{s}}{\text{mL}}$. Visually we can compare the left and right ventricular pressure-volume loops before and after the changes in data for each of our patients in this analysis. Figure S7 shows this comparison for patient 839. Figures S5 and S6 show that patients 456 and 794 have outliers in both the nominal and optimized $\hat{\theta}$ subset. While their respective data did not show any outliers, we studied the parameters which we felt would normalize the parameters to the average. Patient 456 showed a high pulmonary valve resistance and their right ventricular pressure at systole was collected as 44 mmHg. Lowering the pressure to 37 mmHg allowed the pulmonary valve resistance to lower to approximately .013 $\frac{\text{mmHg}\cdot\text{s}}{\text{mL}}$, which in turn allowed patient 456 to be within the average of the rest of the patients. Table S4 shows the comparisons of the data and optimized parameter values for patient 456 and figure S8 shows the pressure-volume loops. Patient 794 showed high pulmonary vein elastance and $\lambda_{lv}$ and their right ventricular pressure at systole was collected as 38 mmHg with a wedge pressue of 21 mmHg. Lowering the pressure to 36 mmHg and lowering the wedge pressure to 16.66 mmHm allowed the pulmonary vein elastance to lower to approximately .349 $\frac{\text{mmHg}}{\text{mL}}$, which in turn allowed patient 794 to be within the average of the rest of the patients for this parameter, however, the $\lambda_{lv}$ parameter remained high. Table S5 shows the comparisons of the data and optimized parameter values for patient 794 and figure S9 shows the pressure-volume loops.

Table S3: Patient 839 first RHC data and nominal parameters. Pulmonary arterial pressure at systole was raised from 18 mmHg to 24.5 mmHg in order to see the effects on the resistance in the pulmonary valve. This increase in pressure allowed the resistance to decrease to $\sim .0049\frac{\text{mmHg}\cdot\text{s}}{\text{mL}}$

| **Data** | $P_{rv,sys}$ | $P_{rv,diast}$ | $P_{pa,sys}$ | $P_{pa,diast}$ | $P_{pcw}$ | $P_{sa,sys}$ | $P_{sa,diast}$ | CO |
|---|---|---|---|---|---|---|---|---|
| **Before** | 25 | 1 | 18 | 9 | 7.33 | 135 | 84 | 101.67 |
| **After** | 25 | 1 | 24.5 | 9 | 7.33 | 135 | 84 | 101.67 |
| **Parameters** ($\times 10^{-2}$) | $\lambda_{lv}$ | $\lambda_{rv}$ | $E_{pa}$ | $E_{pu}$ | $R_{pul}$ | $E_{sa}$ | $R_{sys}$ | $R_{pv}$ |
| **Before** | 2 | 1.5 | 20.37 | 11.93 | 10.5 | 75.74 | 130.94 | 6.89 |
| **After** | 2 | 1.5 | 27.73 | 11.93 | 16.89 | 75.73 | 130.94 | .49 |

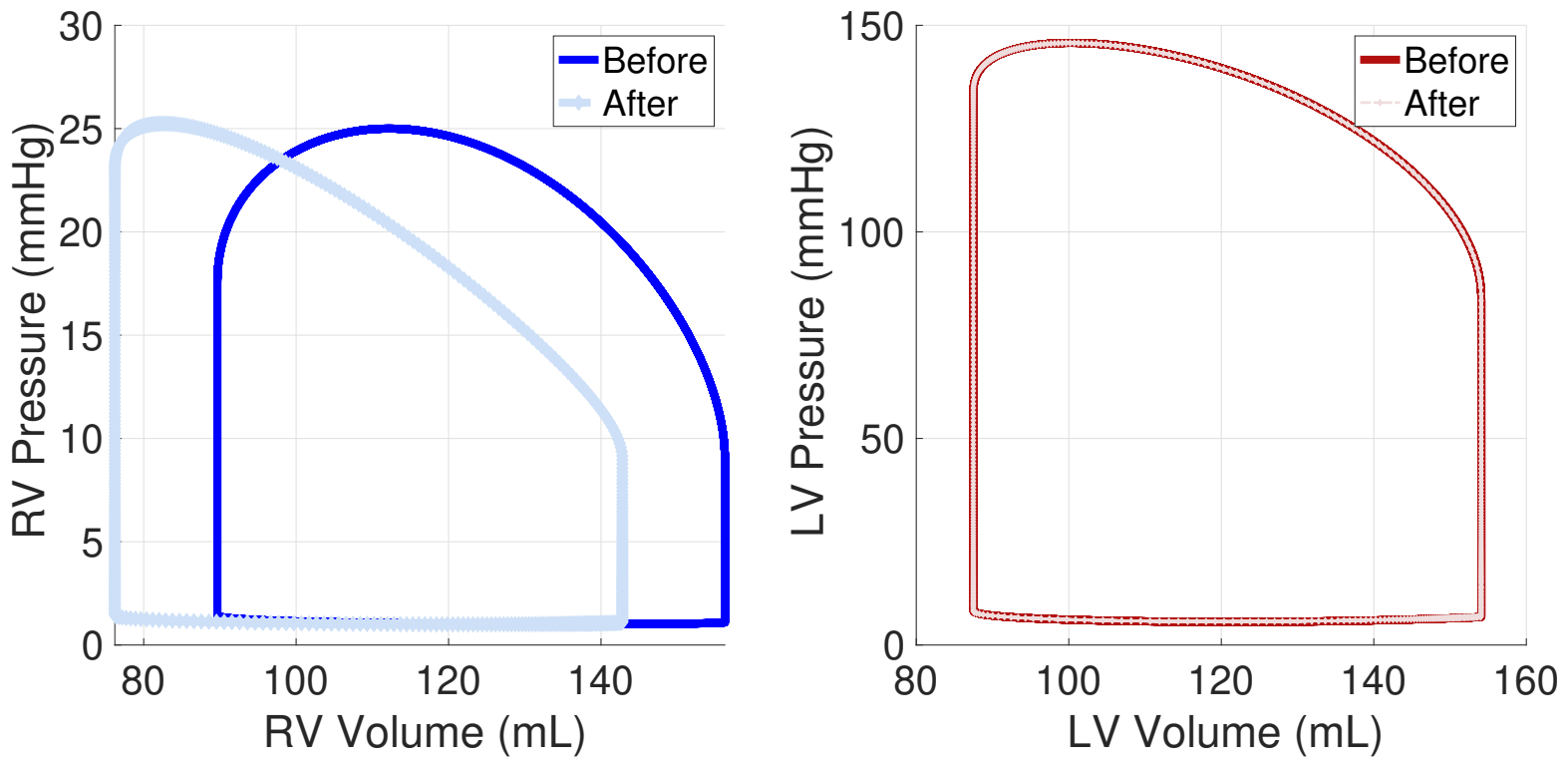

Figure S7: Comparison of the left and right ventricular pressure-volume loops for patient 839 before and after the change in pulmonary arterial pressure at systole.

Table S4: Patient 456 first RHC data and optimized parameters. Right ventricular pressure at systole was lowered from 44 mmHg to 37 mmHg in order to see the effects on the resistance in the pulmonary valve. This decrease in pressure allowed the resistance to decrease by $\sim .013\frac{\text{mmHg}\cdot\text{s}}{\text{mL}}$

| **Data** | $P_{rv,sys}$ | $P_{rv,diast}$ | $P_{pa,sys}$ | $P_{pa,diast}$ | $P_{pcw}$ | $P_{sa,sys}$ | $P_{sa,diast}$ | CO |
|---|---|---|---|---|---|---|---|---|
| **Before** | 44 | 4 | 35 | 13 | 13 | 104 | 67 | 91 |
| **After** | 37 | 4 | 35 | 13 | 13 | 104 | 67 | 91 |
| **Parameters** ($\times10^{-2}$) | $\lambda_{lv}$ | $\lambda_{rv}$ | $E_{pa}$ | $E_{pu}$ | $R_{pul}$ | $E_{sa}$ | $R_{sys}$ | $R_{pv}$ |
| **Before** | 3.29 | 2.17 | 60.6 | 7.48 | 9.15 | 75.7 | 87.8 | 2.6 |
| **After** | 3.29 | 2.27 | 59.35 | 7.21 | 9.19 | 75.69 | 87.77 | 1.32 |

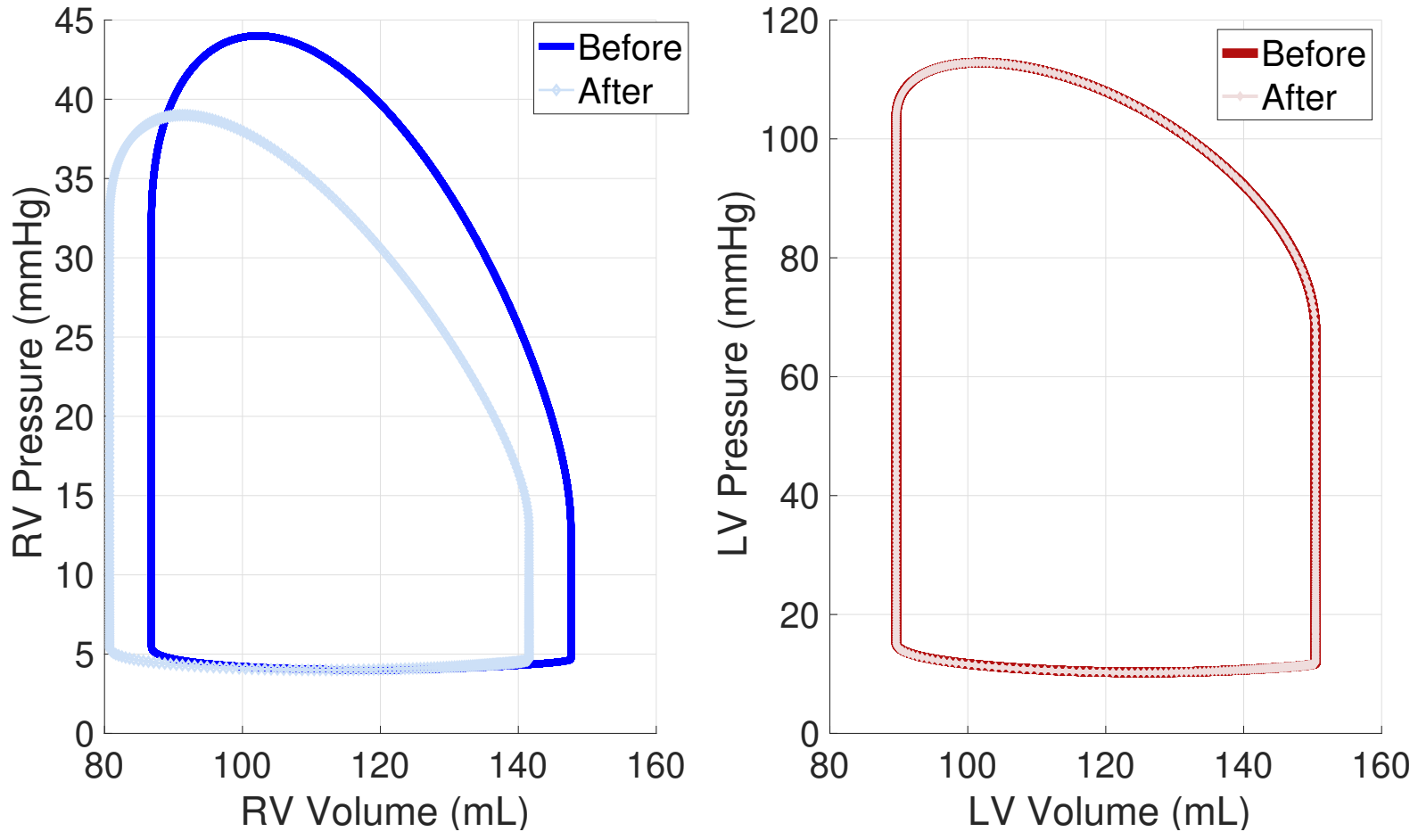

Figure S8: Comparison of the left and right ventricular pressure-volume loops for patient 456 before and after the change in right ventricular pressure at systole. The left ventricular pressure-volume loop remains constant as expected.

Table S5: Patient 794 first RHC data and optimized parameters. Right ventricular pressure at systole was lowered from 38 mmHg to 36 mmHg and the wedge pressure was lowered from 21 mmHg to 16.66 mmHg. Consequently, pulmonary arterial pressure at systole was lowered to ensure blood flow in the right direction. The parameters in question $\lambda_{lv}$ and $E_{es,pu}$ did decrease as expected, however, it did not decrease enough to be within the mean and standard deviation.

| **Data** | $P_{rv,sys}$ | $P_{rv,diast}$ | $P_{pa,sys}$ | $P_{pa,diast}$ | $P_{pcw}$ | $P_{sa,sys}$ | $P_{sa,diast}$ | CO |
|---|---|---|---|---|---|---|---|---|
| **Before** | 38 | 6 | 36 | 17 | 21 | 98 | 67 | 113.8 |
| **After** | 36 | 6 | 34* | 17 | 16.66* | 98 | 57 | 113.8 |
| **Parameters** ($\times10^{-2}$) | $\lambda_{lv}$ | $\lambda_{rv}$ | $E_{pa}$ | $E_{pu}$ | $R_{pul}$ | $E_{sa}$ | $R_{sys}$ | $R_{pv}$ |
| **Before** | 4.48 | 3.16 | 3.49 | 40.06 | 4.72 | 62.5 | 60.34 | .657 |
| **After** | 4.29 | 3.19 | 29.02 | 34.87 | 7.86 | 65.91 | 61.46 | .693 |

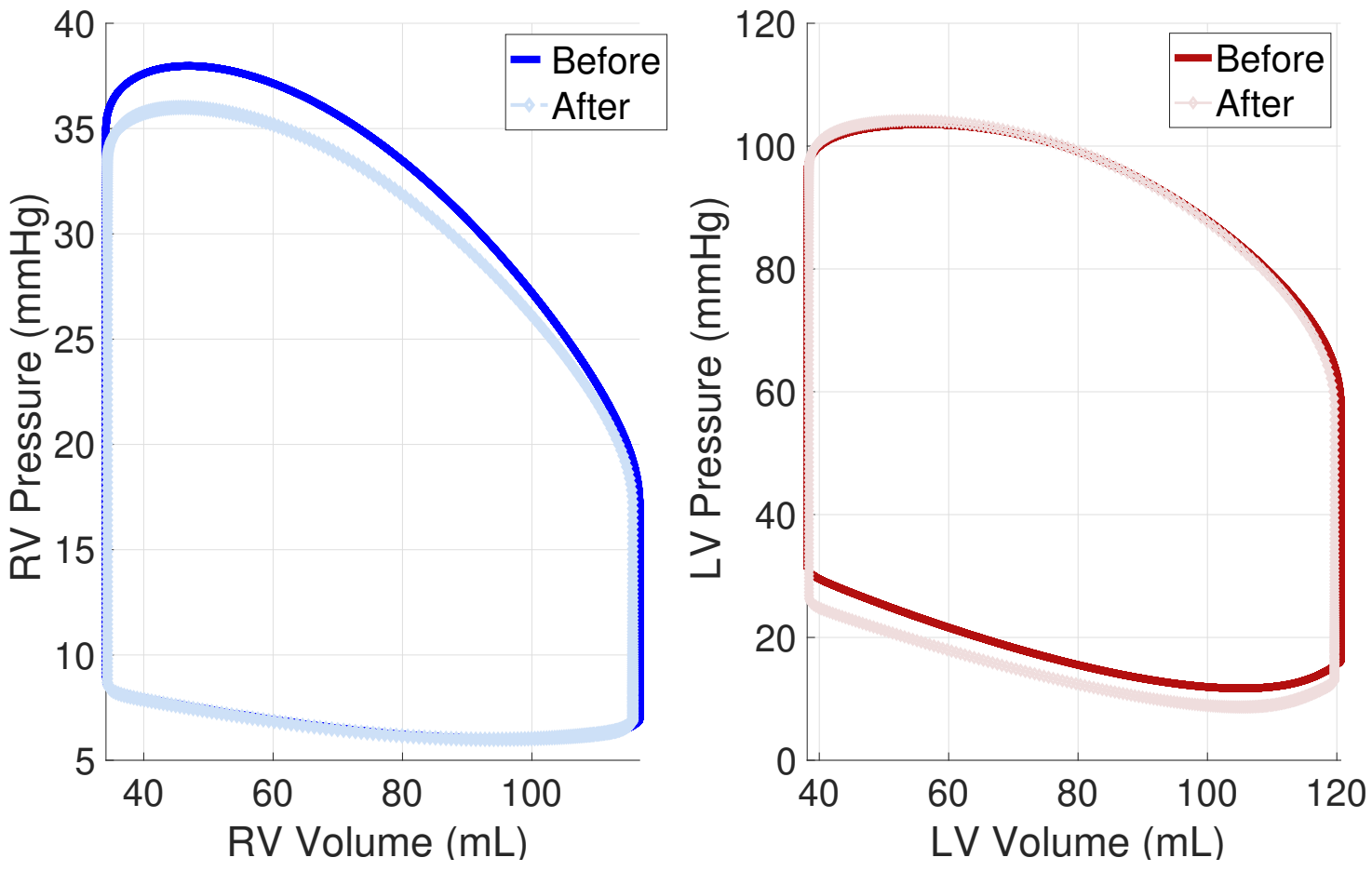

Figure S9: Comparison of the left and right ventricular pressure-volume loops for patient 794 before and after the change in right ventricular pressure at systole.